\documentclass{emulateapj}
\usepackage{natbib}
\usepackage{color}
\usepackage{graphicx}
                                          
\tightenlines

\newcommand     \Bv    {{\bf B}}  
\newcommand     \vv    {\,{\bf v}}  
\newcommand     \kv    {\,{\bf k}}  
\newcommand     \mfp    {{\rm mfp}}  
\newcommand     \A    {\,{\rm A}}  
\newcommand     \res    {{\rm res}}  
\newcommand     \pc    {\,{\rm pc}}  
\newcommand     \cri    {{\rm cri}}  
\newcommand     \cm    {{\rm cm}}  
\newcommand     \eff     {{\rm eff}}  
\newcommand     \s       {{\rm s}}

\newcommand     \K       {{\rm K}}
\def     \H   {{\rm H}}
\def     \cas {{\rm cas}}

\def     \gas  {{\rm gas}}

\def    \yr       {{\rm yr}}
\def    \drag   {{\rm drag}}
\def    \d        {{\rm d}}
\def    \TTD   {{\rm TTD}}

\newcommand      \Zmeant {$\langle Z\rangle$~}
\def	\Angstrom	{\,{\rm \AA}}		
\def \bea {\begin{eqnarray}}
\def \ena {\end{eqnarray}}

\begin{document}
\title{Revisiting Acceleration of Charged Grains in MHD turbulence}
\author{Thiem Hoang \altaffilmark{1}, A. Lazarian \altaffilmark{1}, \& R. Schlickeiser \altaffilmark{2}}

\affiliation{
$^1$Department of Astronomy, University of Wisconsin, Madison, WI 53706, USA\\
$^2$Institut f$\ddot{\rm u}$r Theoretische Physik, Lehrstuhl IV: Weltraum- und 
Astrophysik, Ruhr-Universit$\ddot{\rm a}$t Bochum, 44780 Bochum, Germany}

\begin{abstract}
We study the acceleration of charged grains by magnetohydrodynamics (MHD)
turbulence in the interstellar medium (ISM). We begin with revisiting gyroresonance
acceleration by taking into account the fluctuations of grain guiding center
along a uniform magnetic field (i.e. nonlinear theory--NLT). We calculate grain
velocities due to gyroresonance by fast MHD modes using the NLT for
different phases of the ISM, and compare with results obtained using
quasi-linear theory (QLT). We find for the parameters applicable to the
typical ISM phases that the fluctuations of grain guiding center
reduce grain velocities by less than $15\%$, but they can be important for
more special circumstances. We confirm that large grains can be accelerated
to super-Alfv\'{e}nic velocities through gyroresonance. For such
super-Alfv\'{e}nic grains, we investigate the effect of further acceleration
via transit time damping (TTD) by fast modes. We find that due to the
broadening of resonance condition in the NLT, the TTD acceleration is
not only important for the cosines of grain pitch angle relative
to the magnetic field $\mu>V_{\A}/v$, but also for $\mu<V_{\A}/v$ where
$v$ is the grain velocity and $V_{\A}$ is the Alfv\'{e}n speed.
We show that the TTD acceleration is dominant over the gyroresonance
for large grains, and can increase substantially grain velocities induced
by gyroresonance acceleration. We quantify another stochastic acceleration
mechanism arising from low frequency Alfv\'{e}n waves. We discuss the
range of applicability of the mechanisms and their implications.
\end{abstract}

\keywords{dust, extinction -- ISM: kinematics and dynamics, acceleration -- ISM}

\maketitle

\section{Introduction}

Dust grains play crucial roles in many aspects of the interstellar medium (ISM).
For example, alignment of dust grains with respect to magnetic field provides
insight into star formation  through far-infrared and submm polarized emission
(see Lazarian 2007 for a review).
Very small spinning dust grains radiate microwave emission that contaminate 
to cosmic microwave background (CMB) radiation (Draine \& Lazarian 1998;
Hoang, Draine \& Lazarian 2010; Hoang, Lazarian \& Draine 2011).
Optical extinction and polarization properties of dust depend mainly on its
size distribution. Grain-grain collisions, which depend on grain relative
motions, govern grain coagulation and destruction, that result in the grain
size distribution (see Hirashita \& Yan 2009). Grain-grain collisions are
considered the first stage of planetesimal formation in circumstellar disks
(see e.g., Dullemond \& Dominik 2005).

Traditionally it is believed that the motion of dust grains in the ISM arises
from radiative force, ambipolar diffusion and hydrodrag (see Draine 2011).
The resulting motion from these processes is sub-Alfv\'{e}nic (i.e., $v\ll
V_{\A}=B/\sqrt{4\pi\rho}$ where $B$ is the magnetic field strength and
$\rho$ is the gas mass density), except in special environment conditions
(see Purcell 1969; Roberge et al. 1993).

Astrophysical environments are practically all magnetized and turbulent,
and turbulence is expected to be an important factor in accelerating dust
grains. The evidence for turbulence from electron density fluctuations
testifies for the existence of the so-called Big Power Law in the Sky
(Armstrong et al. 1995; Chepurnov \& Lazarian 2010), while the fluctuations
of velocity (see Lazarian 2009 and references therein) provide convincing
evidences of the dynamic nature of the observed inhomogeneities.
Whether an environment is thermally dominated or magnetized dominated
depends on the plasma $\beta$ parameter, which is defined as the ratio
of gas pressure to magnetic pressure
\bea
\beta=\frac{8\pi n_{\H}k_{\rm B}T_{\gas}}{B^{2}}=0.1\left(\frac{n_{\H}}{30~\cm^{-3}}\right)
\left(\frac{T_{\gas}}{100~\K}\right)\left(\frac{10~\mu G}{B}\right)^{2},~~~
\ena
where $n_{\H}$ is the gas density and $T_{\gas}$ is the gas temperature.

Recent decade has been marked by substantial progress in understanding
of MHD turbulence. This included generalizing incompressible Alfv\'{e}nic
turbulence\footnote{While there are still ongoing debates about the detailed
structure and dynamics of incompressible MHD turbulence, we believe that
Goldreich \& Sridhar (1995) model provides an adequate starting point. In fact,
recent studies in Beresnyak \& Lazarian (2010) and Beresnyak (2011) provided
additional supports for the model.} by Goldreich \& Sridhar (1995) to realistically
compressible media and successful testing of the compressible theory (Lithwick
\& Goldreich 2001; Cho \& Lazarian 2002, 2003; Kowal \& Lazarian 2010). In
what follows in describing compressible MHD turbulence we shall be guided
by the mode decomposition of MHD turbulence into Alfv\'{e}n, slow and, fast
modes presented in Cho \& Lazarian (2002, 2003).

Studies of grain acceleration for magnetized turbulent environments were
initiated by Lazarian \& Yan (2002) who dealt with the acceleration by
incompressible Alfv\'{e}nic turbulence. Comprehensive studies of the acceleration
in realistically compressible environments were performed in Yan \& Lazarian
(2003, hereafter YL03) and Yan, Lazarian \& Draine (2004, hereafter YLD04).
Those studies identified gyroresonant interactions of grains with fast MHD
modes as a new powerful mechanism of grain acceleration. Recently,
Yan (2009) considered betatron acceleration and came to the conclusion
that for most environments the betatron acceleration is subdominant to
the gyroresonance acceleration for sub-Alfv\'{e}nic grains. The application
of grain velocities predicted by gyroresonance for modeling dust extinction
curve provided good correspondences between observations and theoretical
predictions (Hirashita \& Yan 2009) indicating that the turbulence is indeed
the main driving force behind grain acceleration.

The gyroresonance acceleration due to compressible MHD turbulence
(YL03;YLD04) was studied using quasi-linear theory (QLT, Jokipii 1966;
Schlickeiser \& Miller 1998). The underlying assumption of the QLT is that
the guiding center is assumed to move in a regular trajectory along a
uniform magnetic field $\Bv_{0}$. The condition for a grain with velocity $v$
to resonantly interact with fast MHD modes at the scale $k_{\|}$ is given by
$\omega-k_{\|}v\mu-\omega=n\Omega$, for $n=0, \pm 1,\pm2, ...$ where
$\mu$ is the cosine of the grain pitch angle between $\vv$ and $\Bv_{0}$,
$\omega$ is the wave frequency, and $\Omega$ is the Larmor frequency
of the charged grain around $\Bv_{0}$. The gyroresonance acceleration
with $n\ne 0$ is dominant by eddies with size equal to gyro radius $l\sim r_{g}$.

Transit-time damping (TTD) or transit-time acceleration, arises from resonant
interactions of particles with the compressive component of magnetic fluctuations
(i.e., the component parallel to the mean magnetic field $\Bv_{0}$).
When the grain moves together with the wave along $\Bv_{0}$, it is
subject to magnetic mirror forces $-(mv_{\perp}^{2}/2B)\nabla_{\|}\Bv$, where
$m$ is the grain mass, $v_{\perp}$ is the grain velocity component perpendicular
to $\Bv_{0}$, and $\Bv$ is the total magnetic field.
In the plasma reference, the back and forth collisions of the grain with
the moving magnetic mirrors increase grain energy because the head-on
collisions are more frequent than trailing collisions due to the larger relative
velocity between grain and wave (see Fisk 1976; Schlickeiser \& Miller 1998).
The TTD acceleration with resonance condition
$k_{\|}v\mu=\omega$, was disregarded in previous studies on grain acceleration
because grains are expected to move slowly along the uniform magnetic field,
for which they can not catch up with the propagation of magnetic mirrors
along this direction. Although the gyroresonance is found to be able to
accelerate large grains to super-Alfv\'{e}nic velocities (see YL03; YLD04),
which is sufficient to trigger TTD, the resulting grain motion mostly
perpendicular to the uniform field $\Bv_{0}$ (i.e. $v\mu=0$) in the
QLT regime makes TTD incapable.

Due to magnetic fluctuations in the ISM, the local magnetic field $\Bv$ can
be decomposed into a uniform field plus a turbulent component, i.e.,
$\Bv=\Bv_{0}+\delta \Bv$. Thus, any perturbation $\delta \Bv$ will induce
the fluctuations of grain guiding center from a regular trajectory along
the uniform magnetic field (see e.g., Shalchi 2005). Non-linear theory
(hereafter NLT) for gyroresonance that takes into account such fluctuations
of guiding center was formulated in Yan \& Lazarian (2008, hereafter YL08)
to describe the propagation of energetic particles, and later it was applied
in Yan et al. (2008, hereafter YLP08) to study acceleration of energetic
particles in solar flares.

The important modification present in the NLT is the broadening of resonance
function from a Delta function $\delta(\omega-k_{\|}v_{\|}-n\Omega)$ to a
Gaussian function $R_{n}(\omega-k_{\|}v_{\|}-n\Omega)$ (see YL08). Such a
broadening of the resonance condition allows grains moving with $v\sim V_{\A}$
perpendicular to $\Bv_{0}$ to have TTD with compressive waves propagating
along $\Bv_{0}$. We are going to clarify the effects that TTD induces on grain
acceleration in the present paper.

In what follows, we revisit the basics of resonance acceleration for charged
grains by taking into account additional physical processes that were not
considered within original treatments. In discussing the gyroresonance
acceleration, we are going to take into account the fluctuations of grain
guiding center. In particular, we are going to investigate the efficiency of
TTD on grain acceleration in MHD turbulence.

The structure of the paper is as follows. In \S 2, we present briefly the
problem of grain charging, important dynamical timescales, and identify
the range of grain size in which grain charge fluctuations are important.
We revisit gyroresonance acceleration, and introduce TTD acceleration
in \S 3. Grain velocities induced by gyroresonance acceleration and TTD
are presented in \S 4. \S 5 is devoted for stochastic acceleration by low
frequency Alfv\'{e}n waves. Discussion and summary are presented in
\S 6 and 7, respectively.

\section{Grain Charging and Dynamics}
\subsection{Grain Charging}

Charging processes for a dust grain in the ISM consist of its sticking collisions
with charged particles in plasma (Draine \& Sutin 1985) and photo-emission induced
by $h\nu\ge 13.6$ eV photons (Weingartner \& Draine 2001).
In the former case, the grain acquires charge by capturing electrons and ions
from the plasma, while in the latter case the grain looses charge by emitting
photoelectrons. After a sufficient time, these processes result in a statistical
equilibrium of ionization, and the grain has a mean charge, denoted by
$\langle Z\rangle$, which is equal to the charge averaged over time. Due to
the discrete nature of charging events, the grain charge fluctuates around
$\langle Z\rangle$. The probability of finding the grain with charge $Ze$ is
described by charge distribution function $f_{Z}$. Here we find the charge
distribution $f_{Z}$ using statistical ionization equilibrium as in
Draine \& Sutin (1985) and Weingartner \& Draine (2001). Hoang \& Lazarian (2011)
found that the statistical ionization equilibrium is not applicable for tiny grains
with size $a<10\Angstrom$ for which the charging is infrequent, but it
is adequate for grains considered in the present paper.

Figure \ref{timescale} shows the variation of the grain mean charge
$|\langle Z\rangle|$ for graphite and silicate grains in the cold neutral
medium (CNM), warm neutral medium (WNM), warm ionized medium
(WIM). In the WIM, $\langle Z\rangle$ varies rapidly with the grain size,
and change its sign at $a\sim  10^{-6}$ and $\sim 10^{-5}$ cm, marked by
filled circles.

Let us define a characteristic relaxation time of the charge fluctuations,
$\tau_{Z}$, which is equal to the time required for the grain charge to
relax from $Z$ to the equilibrium state (Draine \& Lazarian 1998b):
\bea
\tau_{Z}=\frac{\langle (Z-\langle Z\rangle)^{2}\rangle}{\sum_{Z}f_{Z}J_{tot}(Z)}
\equiv\frac{\sigma_{Z}^{2}}{\sum_{Z}f_{Z}J_{\rm tot}(Z)},
\ena
where $J_{\rm tot}(Z)$ is the total charging rate due to collisional charging
and photoemission (see Draine \& Sutin 1987; Weingartner \& Draine 2001).
Here we averaged over all possible charge states $Z$ to find $\tau_{Z}$.

In an ambient magnetic field $\Bv$, the grain with mean charge $\langle Z\rangle e$
gyrates about $\Bv$ on a timescale equal to the Larmor period:
\bea
\tau_{L}=\frac{2\pi m c}{|\langle{Z}\rangle| eB}=1.56\times 10^{2}\left(\frac{a}{10^{-6}~\cm}\right)\left(\frac{\mu G}{|\langle Z\rangle|B}\right) \yr,~~~ \label{tauL}
\ena
where $m=4/3\pi a^{3}\rho_{d}$ with $\rho_{d}$ being the dust mass density is 
the grain mass. We adopt $\rho_{d}=2.2$ and $3.0$ g $\cm^{-3}$ for graphite
and silicate grains, respectively. The Larmor frequency reads
$\Omega=(\langle{Z}\rangle eB)/mc$.

We calculate the relaxation time of charge fluctuations $\tau_{Z}$ for both
graphite and silicate grains in various phases of the ISM with physical
parameters listed in Table 1. Figure \ref{timescale} compares $\tau_{Z}$
with the gas drag time $\tau_{\drag}$ (see Eq. \ref{tau_drag}) and the
Larmor period $\tau_{L}$. It can be seen that $\tau_{Z}\ll \tau_{L}<\tau_{\drag}$
for grains larger than $\sim 2\times 10^{-7}$ cm. For grains smaller than
$\sim2\times 10^{-7}$ cm, $\tau_{Z} \ge \tau_{L}$, so that the assumption
for grains to have a constant charge is no longer valid. As a result, the
fluctuations of grain charge should be accounted for in the treatment of
resonance acceleration for such very small grains. This issue will be
addressed in our future paper, in which we employ Monte Carlo method to
simulate grain charge fluctuations (see e.g., Hoang \& Lazarian 2011). In the
present paper, for the sake of simplicity, we adopt $\langle Z\rangle e$
for grain charge within the entire range of the grain size distribution.

\begin{figure}
\includegraphics[width=0.45\textwidth]{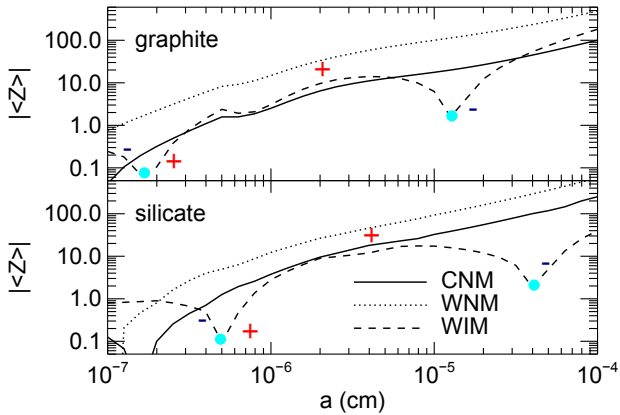}
\caption{Mean grain charge $|\langle Z\rangle|$  as functions of grain size
$a$ for graphite and silicate grains in different ISM phases. For the WIM,
$|\langle Z\rangle|$ changes rapidly with $a$, and filled circles mark
the change in grain charge between being positively charged (+) and
negatively charged (-). }
\label{sigmaZ}
\end{figure}

\begin{figure}
\includegraphics[width=0.45\textwidth]{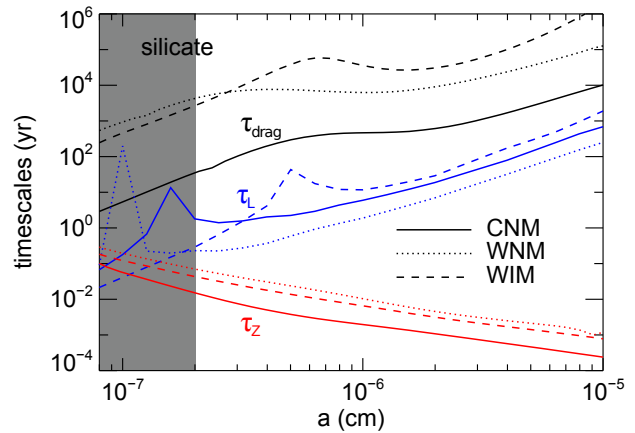}
\caption{Timescales for gas drag $\tau_{\drag}$, Larmor period $\tau_{L}$,
and charge fluctuations $\tau_{Z}$ as functions of the grain size $a$ for
subsonic silicate grains in the various phases of the ISM. Shaded area
marks the range of grain size in which the charge fluctuations are important,
i.e., $\tau_{Z}\geq \tau_{L}$. The peaks in $\tau_{L}$ correspond to
the change in sign of grain charge.}
\label{timescale}
\end{figure}

\subsection{Grain Translational Damping}

Interactions of dust grains with the ambient gas present the primary mechanism
of dissipating translational motions of grains. The damping rate of translational
motion arising from the interaction with neutral gas is essentially the inverse
time for collisions with the mass of the gas equal that of a grain (Purcell 1969),
\bea
\tau_{dn}^{-1}&=&\sqrt{\frac8{\pi}}\frac{n_n}{a\rho_d}(m_{n}k_{\rm B}T_{n})^{1/2},\nonumber\\
&=&2.4\times10^{-12}\left(\frac{10^{-6}\cm}{a}\right)\left(\frac{n_{n}}{30\cm^{-3}}\right)
\left(\frac{T_n}{100~\K}\right)^{1/2}\s^{-1},~~~~
\label{tau_fr}
\ena
where $m_n$, $n_n$, and $T_n$ are the mass, volume density, and temperature
of neutrals, and $a$ is the grain radius.

When the ionization degree is sufficiently high, the interaction of charged
grains with the plasma becomes important. The ion-grain cross section due
to long-range Coulomb forces is larger than the atom-grain cross section. As
a result, the rate of translational motion damping gets modified. For subsonic
motions the effective damping time due to gas drag is renormalized:
\bea
\tau_{\drag}=\alpha^{-1}\tau_{dn} \label{tau_drag}
\ena
with the following renormalizing factor (Draine \& Salpeter 1979)
\bea
\alpha=1+\frac{n_{\rm H}}{2n_n}\sum_{i}x_i\left(\frac{m_{i}}{m_n}\right)^{1/2}
\sum_{Z}f_{Z}\left(\frac{Ze^{2}}{ak_{\rm B}T_i}
\right)^{2}\nonumber\\
\times\ln\left[\frac{3}{2|Z|\sqrt{\pi xn_{\H}}}\left(\frac{k_{\rm B}T_i}{e^2}\right)^{3/2}\right]
.\hspace{.5cm}\label{alpha1}
\ena
Here $x_{i}$ is the abundance of ion $i$ (relative to hydrogen) with mass
$m_{i}$ and temperature $T_i$, $x=\sum_i x_i$, $Ze$ is the grain charge, and
$f_{Z}(Z)$ is the grain charge distribution function. When the grain velocity $v_d$
relative to gas becomes supersonic, the dust-plasma interaction is
diminished, and the damping rate in this case is renormalized due to the
gas-dynamic correction (Purcell 1969),
\begin{equation}
\alpha=\left(1+\frac{9\pi}{128}\frac{v_d^2}{C_{\rm s}^2}\right)^{1/2},\label{alpha2}
\end{equation}
where $C_{\rm s}=\sqrt{k_{\rm B}T_{n}/m_n}$ is the sound speed.

If $\tau_{\rm L}$ is greater than $\tau_{\drag}$, then the effect of magnetic
field on dust dynamics is negligible. However, $\tau_{\rm L} \ll \tau_{\drag}$
in most phases of the ISM.

\begin{table}
\caption{Idealized Environments and MHD turbulence parameters}\label{ISM}
\begin{tabular}{llll} \hline\hline\\
\multicolumn{1}{c}{\it Parameters} & \multicolumn{1}{c}{CNM}& 
{WNM} &WIM\\[1mm]
\hline\\
$n_{\rm H}$~(cm$^{-3}$) &30 &0.4 &0.1 \\[1mm]
$T_{\rm gas}$~(K)& 100 & 6000 &8000\\[1mm]
$x_{\rm H}$ &0.0012 &0.1 &0.99 \\[1mm]
$B~(\mu G)$ &6 &5.8 &3.35 \\[1mm]
$L~(\pc)$ &0.64 &100 &100 \\[1mm]
$\delta V=V_{\A}~({\rm km}~\s^{-1})$&{2} &20 &20 \\[1mm]
$k_{c}~({\cm}^{-1})$&{$7\times 10^{-15}$} &$4\times 10^{-17}$ &... \\[1mm]
{Damping}&{Neutral-ion} &{Neutral-ion} &{Ion~ viscous}\\[1mm]
& & & {and~ collisionless} \\[1mm]
\hline
\footnotetext{Here $n_{\H}$ is the gas density, $T_{\gas}$ is the gas temperature,
$x_{\H}$ is the ionization fraction of H, $B$ is the strength of magnetic field,
$L$ is the injection scale of turbulence, $\delta V$ is the rms velocity of turbulence
at the injection scale, and $k_{c}$ is the cutoff scale of turbulence due to
collisional and collisionless damping.}
\end{tabular}
\end{table}

\section{Resonance Acceleration}
\subsection{Gyroresonance acceleration: nonlinear theory}

In this section, we revisit the treatment of resonance acceleration by
fast modes in compressible MHD turbulence using nonlinear theory (NLT).

Consider a grain of mass $m$, charge $Ze$, moving with velocity $v$ in a
magnetized turbulent medium with a uniform magnetic field $\Bv=\Bv_{0}$.
The motion of such charged grain in $\Bv$ consists of the gyration of the
grain about its guiding center and the translation of the guiding center
along $\Bv$. In the QLT limit, the guiding center is assumed to follow a
regular trajectory along $\Bv$ with a constant cosine of pitch angle $\mu=\cos\beta$
with $\beta$ being the angle between $\vv$ and $\Bv$. Gyroresonant
interactions between grain and wave occur when the wave
frequency in a reference system fixed to the grain guiding center is
a multiple of the Larmor frequency:
\bea
\omega-k_{\|}v\mu=n\Omega \label{eq:gyro_res}
\ena
with $n=\pm 1, \pm2, ...$. This resonance condition is equivalently described by
a Delta function $\delta_{n}(\omega-k_{\|}v\mu-n\Omega)$.

Gyroresonance accelerates grains in the direction perpendicular
to the mean magnetic field $\Bv_{0}$ because electric field induced by plasma
perturbations is perpendicular to $\Bv_{0}$ (see e.g. YLD04). This acceleration
mechanism is dominant by eddies smaller than the grain gyroradius,
i.e., $l\le r_{g}$. Indeed, consider gyroresonance by fast MHD modes in
low-$\beta$ plasma. From the resonance condition (\ref{eq:gyro_res}) for
$n=1$ with $\omega=k_{\|}v_{\A}$, we obtain turbulent scales for gyroresonance
$k\ge k_{\res}=r_{g}^{-1}$ or $l\le r_{g}$, where the fact that $\mu\ge -1$ has been used.

In a turbulent medium, the local magnetic field $\Bv=\Bv_{0}+\delta \Bv$
where $\delta \Bv$ is the turbulent component of magnetic field, varies both
in space and time, so $\mu$ changes, and $v_{\|}$ and $v_{\perp}$ change
accordingly. The grain guiding center has fluctuations from its regular trajectory
along $\Bv_{0}$. The NLT takes into account such fluctuations of the guiding
center.

Assuming that the projection of the fluctuations of grain guiding center onto
the mean field $\Bv_{0}$ can be described by a Gaussian distribution,
the resonance condition becomes
\bea
R_{n}\left(\omega-k_{\|}v\mu-n\Omega\right)=\frac{\sqrt{\pi}}{k_{\|}\Delta 
v_{\|}}{\rm exp}\left[-\frac{(k_{\|}v\mu-\omega+ n\Omega)^{2}}{k_{\|}^{2}
(\Delta v_{\|})^{2}}\right],~~~
\ena
where $n=0$ and $\pm 1$, and $\Delta v_{\|}$ is the dispersion of
velocity (see YL08 and YLP08; Appendix C).

We are interested in the grain acceleration, so the diffusion coefficient
arising from gyro-phase averaging $D_{pp}$ is used (see Schlickeiser
\& Miller 1998). In compressible MHD turbulence, the fast modes are shown
to be dominant in gyroresonance acceleration (YL03). Its corresponding
diffusion coefficient is given by (see Appendix C)
\bea
D_{pp}(\mu,p)^{\rm G}=\frac{v\sqrt{\pi}\Omega^{2}(1-\mu^{2})m^{2}V_{\A}^{2}M_{\A}^{2}}
{4LR^{2}}\int_{1}^{k_{c}L}x^{-5/2}dx\nonumber\\
\times\int_{0}^{1}\frac{d\eta}{\eta\Delta \mu}[J_{0}^{2}(w)+J_{2}^{2}(w)]
{\rm exp}\left[-\frac{(\mu-\frac{V_{A}}{\eta v}+n\frac{1}{\eta xR})^{2}}
{(\Delta \mu)^{2}}\right],~~~~\label{eq:dpg}
\ena
where $n=\pm 1$.  In the above equation, $L$ is the injection scale of
turbulence, $w={k_{\perp} v_{\perp}}/{\Omega}, x=k/k_{\min}=kL,
R=vk_{\min}/\Omega, M_{\A}^{2}=
\delta V^{2}/V_{\A}^{2}$, $\eta=\cos\theta$ with $\theta$ is the angle between
the wave vector ${\bf k}$ and the mean magnetic field, $k_{c}$ is the cut-off
of turbulence cascade due to damping, and $J_{n}$ is second order 
Bessel function. The dispersion of the cosine of pitch angle $\Delta\mu$ is
given by Equation (\ref{eq:dmu}) in Appendix C.

\subsection{Transit-Time Damping (TTD)}

Transit-time damping (TTD) or transit-time acceleration, arises from
resonant interactions of particles with the compressive component of
magnetic fluctuations, i.e., the component parallel to the mean
magnetic field $\Bv_{0}$ in magnetized turbulent environments.
When the grain moves together with the wave along $\Bv_{0}$,
it is subject to magnetic mirror forces $-(mv_{\perp}^{2}/2B)\nabla_{\|}\Bv$,
where $v_{\perp}$ is the grain velocity component perpendicular to $\Bv_{0}$.
In the plasma reference, the back and forth collisions of the grain with
the moving magnetic mirrors increase grain energy because the
head-on collisions are more frequent than trailing collisions
(see e.g., Fisk 1976). The resonance condition for a
grain with velocity $\vv$ reads
\bea
\omega-k_{\|}v_{\|}=0,\label{eq:resTTD}
\ena
where $v_{\|}=v\mu$ is the grain velocity component parallel to $\Bv$,
and $\omega$ is the wave frequency (see Fisk 1976; Schlickeiser \& Miller 1998).

For fast MHD modes in low-$\beta$ plasma, the dispersion relation
is $\omega=k V_{\A}$ (see Cho et al. 2002), and the required velocity for
TTD corresponds to $v_{\|}=v_{\A}/\cos\theta$. Thus, if $v_{\|}\ge V_{\A}$,
TTD can be efficient to accelerate grains to large velocities.

In the NLT limit, the diffusion coefficient for TTD is given by (see Appendix C)
\bea
D_{pp}(\mu,p)^{\rm TTD}&=&\frac{v\sqrt{\pi}\Omega^{2}(1-\mu^{2})m^{2}V_{\A}^{2}
M_{\A}^{2}}{2LR^{2}}\int_{1}^{k_{c}L}x^{-5/2}dx\nonumber\\
&&\times\int_{0}^{1}\frac{d\eta}{\eta\Delta \mu}J_{1}^{2}(w){\rm exp}\left[-\frac{(\mu-\frac{V_{\A}}
{\eta v})^{2}}{(\Delta \mu)^{2}}\right].~~~~~~\label{eq:dpt}
\ena

\begin{figure}
\includegraphics[width=0.45\textwidth]{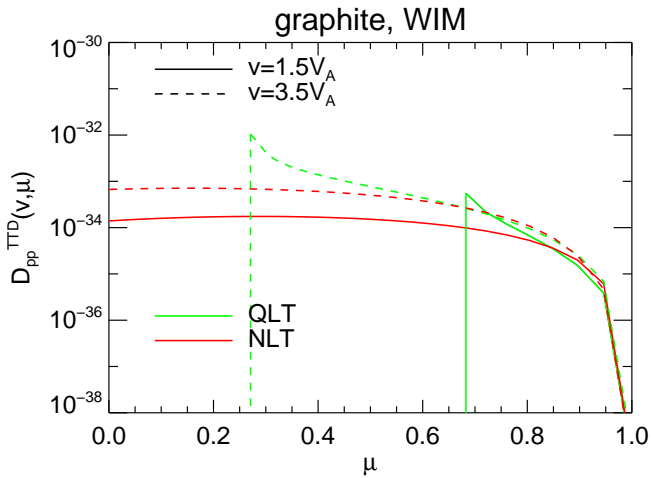}
\caption{Diffusion coefficient $D_{pp}^{\TTD}$ as a function of the cosine
of the grain pitch angle $\mu$ for the limit of QLT (green lines) and NLT (red lines).
Two velocity values of super-Alfv\'{e}nic graphite grains of size $a=10^{-5}$ cm
are considered. $D_{pp}^{\TTD}$ drops sharply for $\mu<V_{\A}/v$ in the QLT,
but $D_{pp}^{\TTD}$ is finite as $\mu \rightarrow 0$ in the NLT as a result of
broadening of resonance conditions.}
\label{Dpp_mu_TTD}
\end{figure}

In Figure \ref{Dpp_mu_TTD} we present $D_{pp}^{\TTD}$ as a function of the
cosine of the grain pitch angle $\mu$ obtained using the QLT and NLT.
Two values of the grain velocity $v=1.5~V_{\A}$ and $3.5~V_{\A}$ are considered.
It can be seen that in the former case, $D_{pp}^{\TTD}$ increases with decreasing
$\mu$ until $\mu=v/V_{\A}$, and drops sharply to zero for $\mu<v/V_{\A}$
because the resonance condition (\ref{eq:resTTD}) is not satisfied.
In contrast, the broadening of the resonance condition in the latter case allows grains
with $\mu<V_{\A}/v$ to have resonant interactions with waves, resulting in finite
$D_{pp}^{\TTD}$ even at $\mu=0$. Therefore, TTD is usually neglected in
the QLT because the gyroresonance tends to accelerate grains
in the perpendicular direction to $\Bv_{0}$, resulting in $\mu=0$, for which
$D_{pp}^{\TTD}\rightarrow 0$. However, TTD can play an important role
in driving grain motion when the fluctuations of guiding center are taken
into account.

\section{Grain Velocities due to Resonance Acceleration}
\subsection{Grain dynamics}
Consider an ensemble of grains with the same mass $m$, moving in the uniform
magnetic field with their different pitch angles $\mu$.
From the equation of motion, $mdv/dt=-mv/t_{\drag}+R$, where $R$ is the
random force, we can obtain
\bea
m\frac{d\langle v^{2}\rangle}{dt}=-\frac{m\langle v^{2}\rangle}{t_{\drag}}+A(v),
\label{eq:dv-dt}
\ena 
where $\langle v^{2}\rangle$ is the grain velocity dispersion averaged over the 
ensemble of grains, $A(v)$ is the rate of energy gain (see YL03).

When the scattering is less efficient than the acceleration,\footnote{The efficiency of scattering relative to acceleration is described by the ratio $p^{2}D_{\mu \mu}/D_{pp}=\cos\theta^{-2}\left({v\cos\theta}/{V_{\A}}+\mu\right)^{2}$
where $\cos\theta=k_{\|}/k$ with $\theta$ is the angle between $\kv$ and $\Bv$ (see Appendix D).
The scattering is negligible for grains with $v< V_{\A}$, but it becomes important
for $v \ge V_{\A}$ or when the grain is moving along the magnetic field, i.e., $v_{\|}\gg v_{\perp}$. } the cosine of the grain pitch angle $\mu=\cos\beta$ changes
slowly during acceleration, and $A(v)$ is given by
\bea
A(v)=\frac{1}{4p^{2}}\frac{\partial}{\partial p}\left(vp^{2}D_{pp}(p,\mu)\right),\label{eq:Av}
\ena
where $D_{pp}(p,\mu)$ is the diffusion coefficient.

When the scattering is more efficient than the acceleration, the pitch angle
can be rapidly redistributed through pitch angle diffusion during the acceleration
(i.e., diffusion approximation). The rate of energy gain (Eq. \ref{eq:Av}) is then
determined by the averaged value $D_{p}$ of $D_{pp}(p,\mu)$ over the
isotropic distribution of $\mu$ (see Dung \& Schlickeiser 1990ab):
\bea
D_{p}(p)=\frac{1}{2}\int_{-1}^{1}\left(D_{pp}(p,\mu)-\frac{D_{\mu p}^{2}(p,\mu)}{D_{\mu\mu}(p,\mu)}\right)d\mu.\label{eq:dp}
\ena
We consider in the present paper zero helicity turbulence with $D_{\mu p}=0$.
When $D_{pp}$ is known, we solve Equation (\ref{eq:dv-dt}) iteratively to get
convergent velocities.

Physical parameters for ISM conditions are shown in Table 1. MHD turbulence 
is injected at a large outer scale $L$ with velocity dispersion $\delta V$.
The value $\delta V$ is chosen such that the turbulence is weak and sub-Alfv\'{e}nic.
Major damping processes are also listed in Table 1. For the CNM and WNM,
the dominant damping arises from the neutral-ion viscosity. The value of $k_{c}$
for these phases is adopted from YLD04. For the WIM, the turbulent damping
arises mainly from ion viscosity and collisionless damping. We calculate $k_{c}$
by equating the damping rate to the rate of turbulence cascade
(see Appendix C, also YL03; YLD04).

Figure \ref{spectrum} sketches possible acceleration mechanisms present
in MHD turbulence. Working scales in the inertial range, spanning from
the injection scale $k_{\min}$ to the cutoff scale $k_{c}$ of turbulence, are indicated.
Gyroresonance works only if $k_{\res}\le k_{c}$, and is dominant at small scales
$k\ge k_{\res}\sim r_{g}^{-1}$. The critical size $a_{\cri}$ for gyroresonance is
then obtained by solving the equation $k_{\res}=k_{c}$ for grain size.

\begin{figure}
\includegraphics[width=0.45\textwidth]{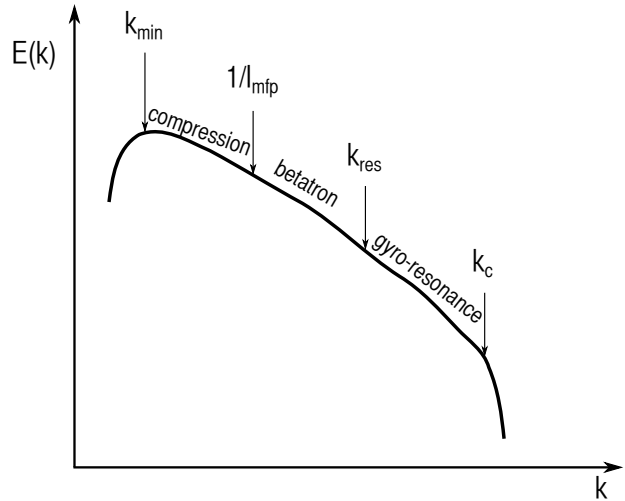}
\caption{A sketch of spectrum of MHD turbulence and acceleration mechanisms
for charged grains are shown with its corresponding scale. $k_{\min}\sim L^{-1}$ is
the injection scale, $l_{\mfp}$ is the grain mean free path, $k_{\res}\sim r_{g}^{-1}$ is the
gyroresonance scale, and $k_{c}$ is the damping cut-off of turbulence.
We assume that grains of interest are large enough so that $k_{\res}<k_{c}$
or Larmor period larger than eddy turnover time.}
\label{spectrum}
\end{figure}

\begin{figure}
\includegraphics[width=0.45\textwidth]{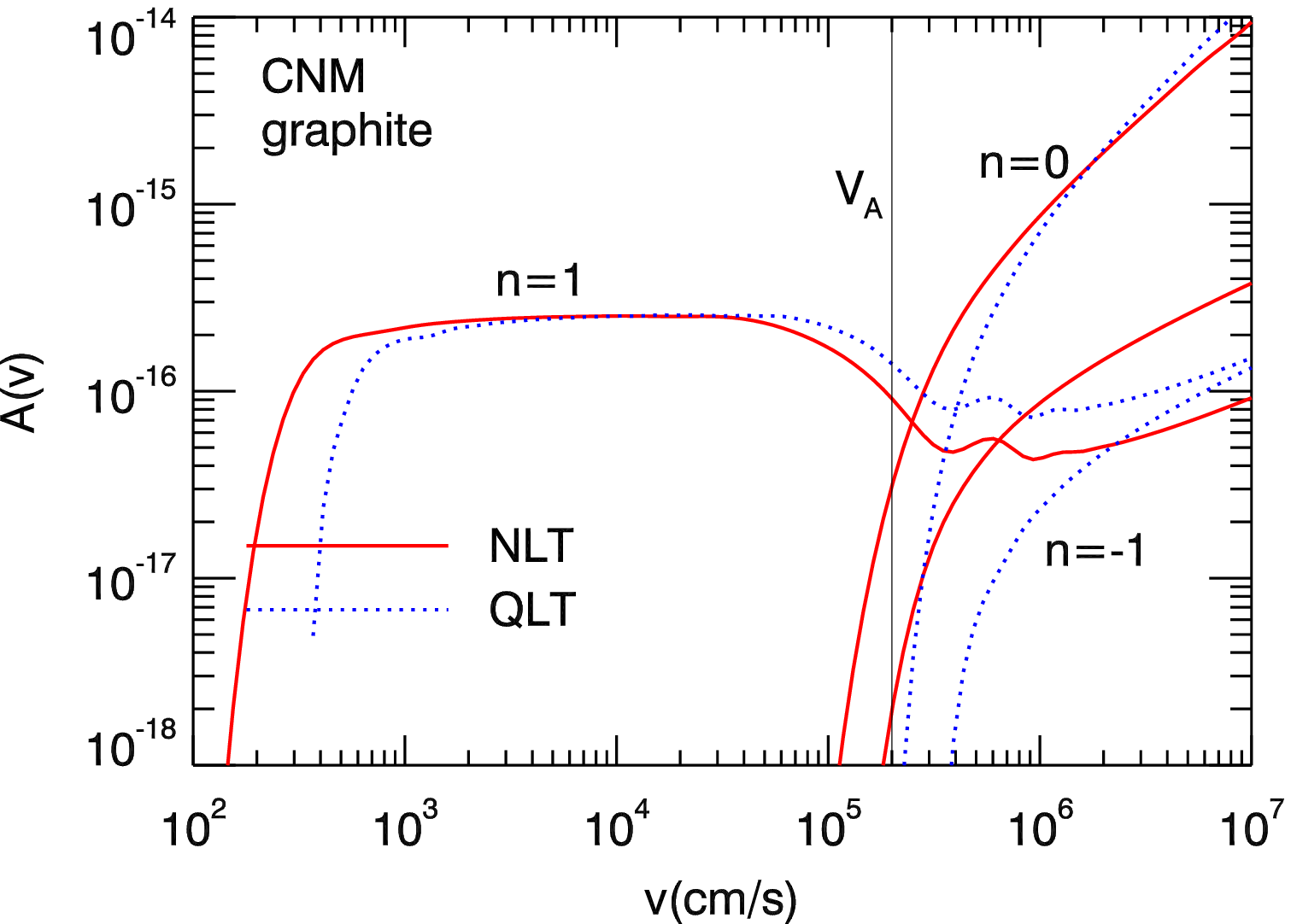}
\includegraphics[width=0.45\textwidth]{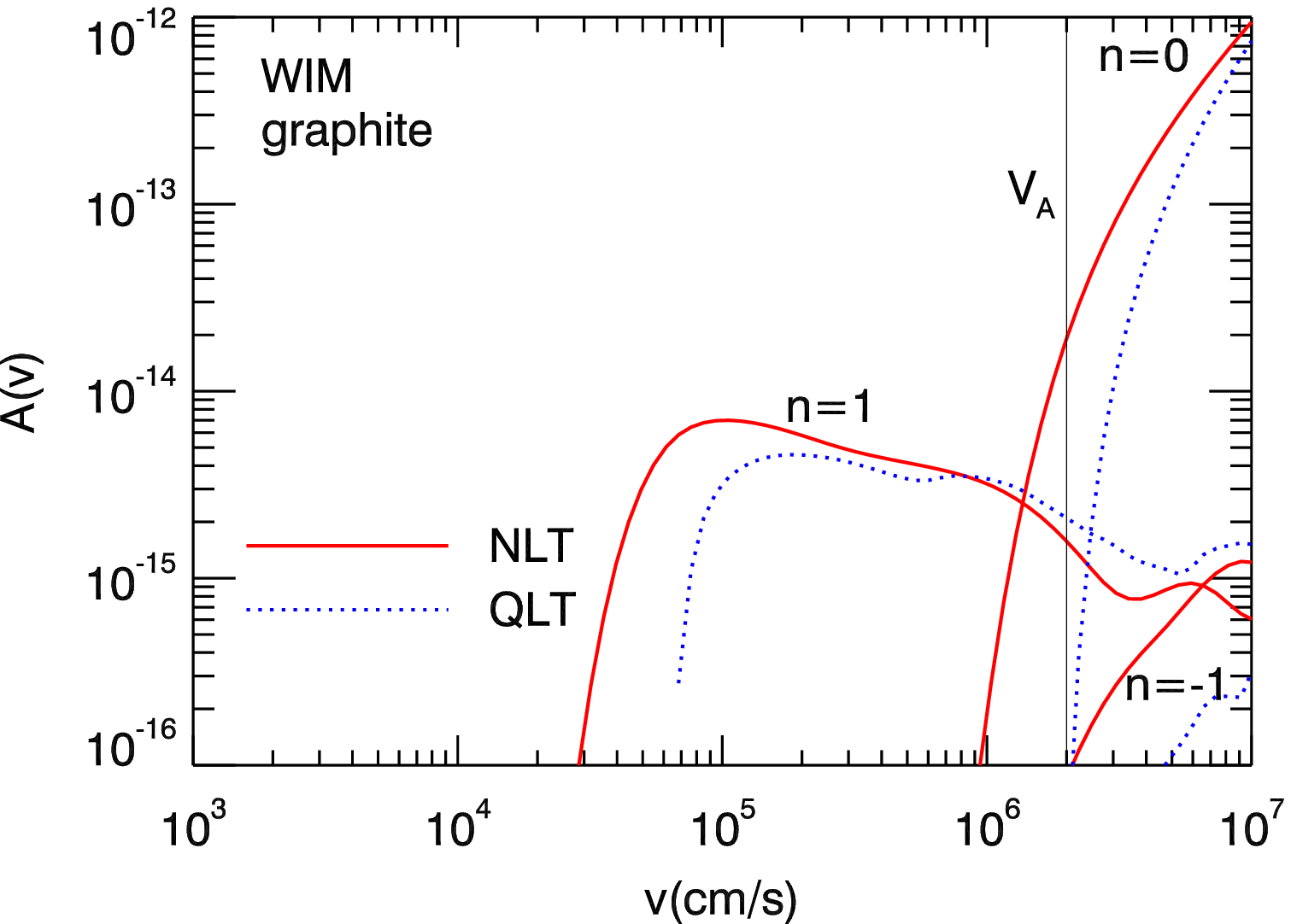}
\caption{Rate of energy gain as a function of the grain velocity for gyroresonance
acceleration ($n=\pm 1$) and transit time damping acceleration ($n=0$, TTD)
for a graphite grain of size $a=10^{-5}$ cm in the CNM ({\it upper}) and WIM
({\it lower}). Solid and dot lines denote results from NLT and QLT, respectively.
The Alfv\'{e}nic speed $V_{\A}$ is indicated. The gyroresonance acceleration ($n=1$)
is dominant for $v< V_{\A}$, and the TTD acceleration ($n=0$) takes over
when $v\ge V_{\A}$. The case of efficient pitch angle scattering is considered.}
\label{Av}
\end{figure}

Figure \ref{Av} shows the rate of energy gain (Eq. \ref{eq:Av}) as a function
of grain velocity $v $ for gyroresonance acceleration ($n=\pm 1$) and TTD
acceleration ($n=0$) in the CNM ({\it upper panel}) and WIM ({\it lower panel})
arising from fast modes in MHD turbulence. Here we assumed that the pitch
angle scattering is efficient. Due to the broadening of resonance condition,
gyroresonant interactions occur at lower velocities in the NLT than QLT
({\it solid lines}). It can also be seen that the gyroresonance acceleration
($n=1$) is dominant  for $v<V_{\A}$, while TTD acceleration
($n=0$) becomes dominant for $v\ge V_{\A}$.

In both the CNM and WIM, as $v\rightarrow V_{\A}$, $A_{v}(n=1)$ becomes
smaller in the NLT than the QLT. As a result, we expect that gyroresonance
acceleration is less efficient in the former case. We are going to quantify such
a difference in the following.

\subsection{Gyroresonance in quasi-linear theory and nonlinear theory}

Grain velocities due to gyroresonance acceleration are obtained by solving
Equation (\ref{eq:dv-dt}) using the diffusion coefficients from Equations (\ref{eq:dpg})
and (\ref{eq:dp}). We assume that at the beginning the grain has low velocity,
so that the pitch angle scattering by TTD is negligible. The gyroresonance
increases rapidly $v_{\perp}$, and $\mu$ decreases to $\mu=0$. 
So, we can assume $\mu=0$ for the gyroresonance acceleration.

\begin{figure}
\includegraphics[width=0.45\textwidth]{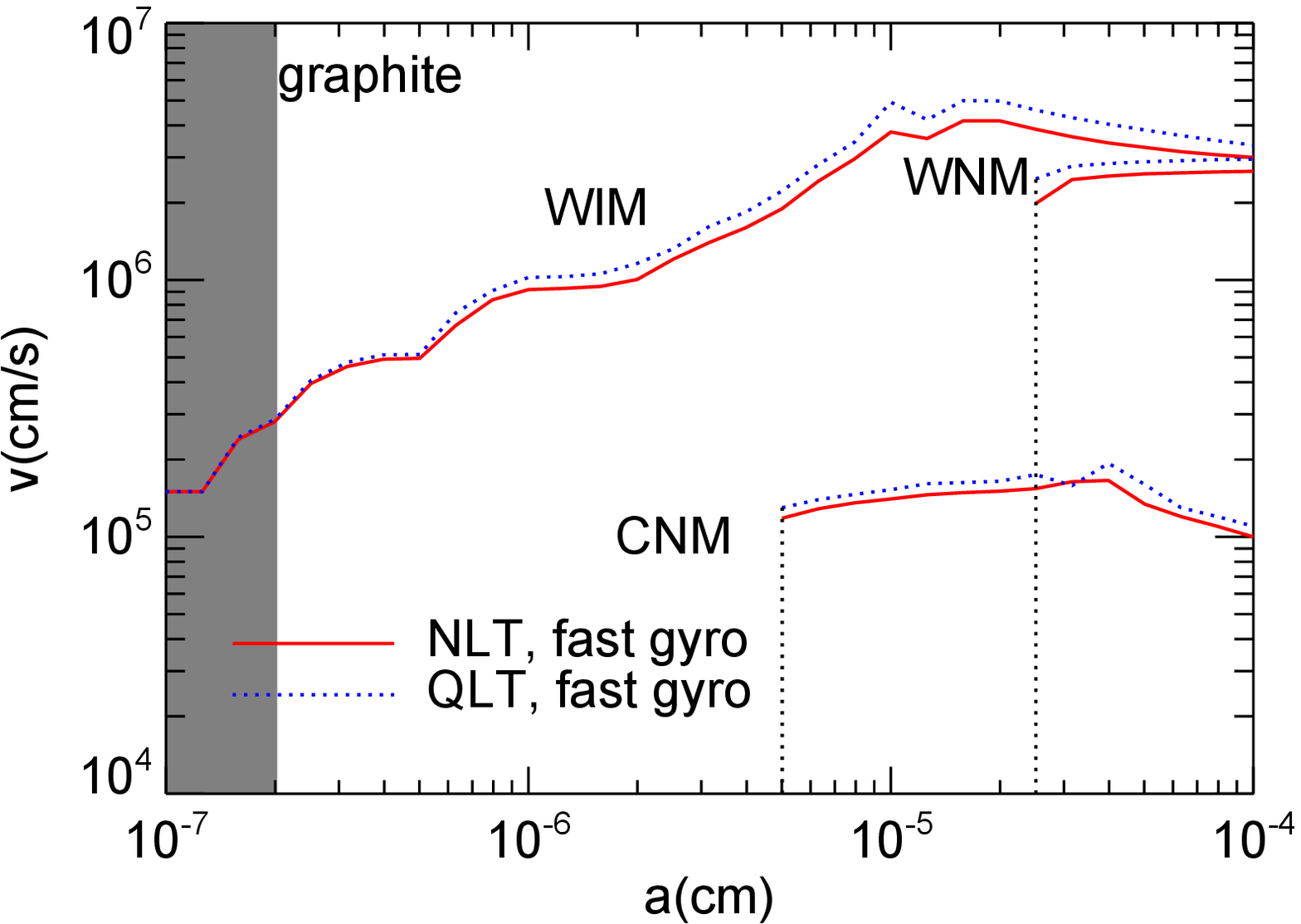}
\includegraphics[width=0.45\textwidth]{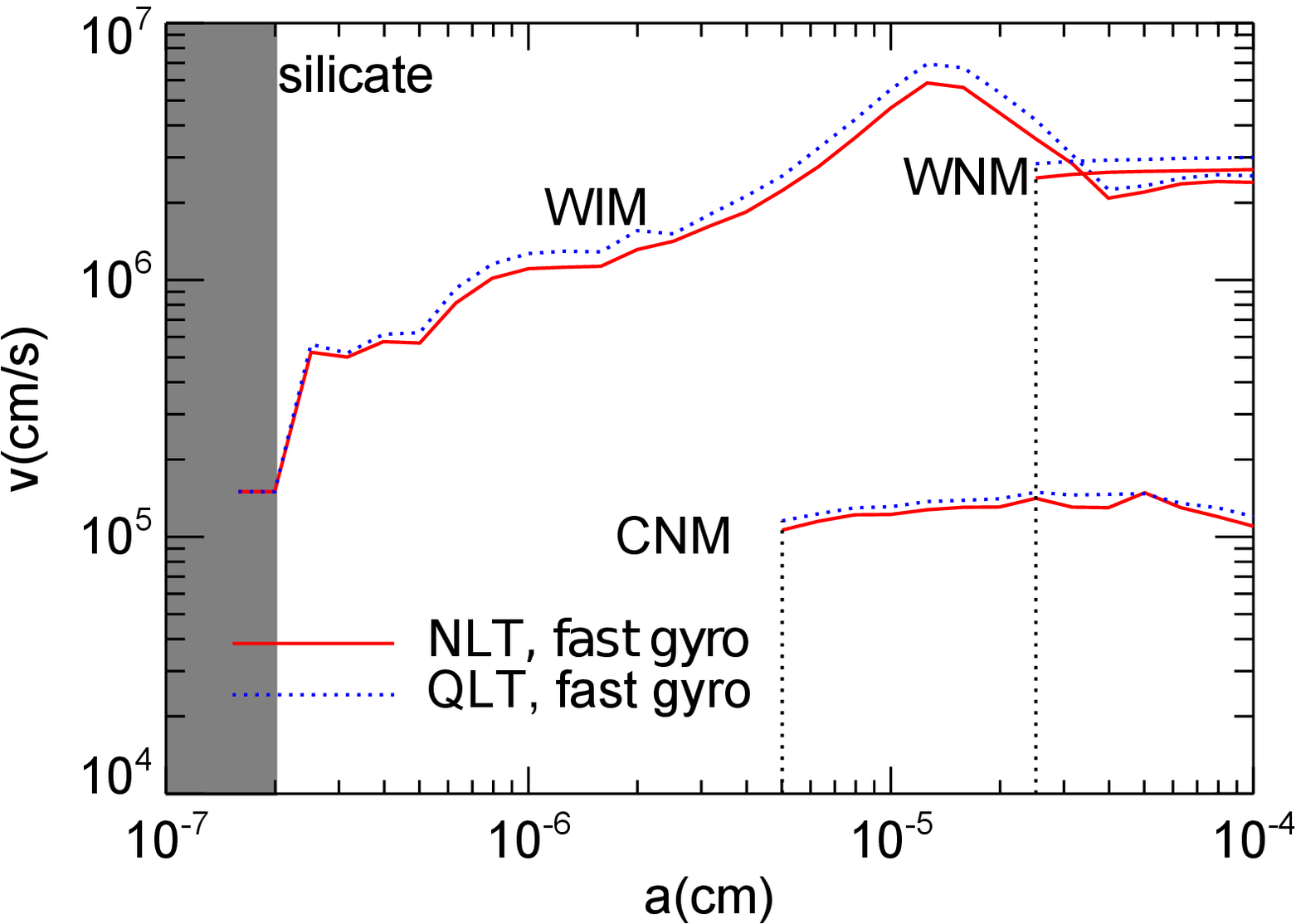}
\caption{Grain velocities relative to gas as a function of grain size for graphite
and silicate grains in various phases of the ISM. Acceleration arising from
gyroresonant interactions of fast modes with grains are obtained using the
QLT ({\it dotted line}) and NLT ({\it solid line}). The difference in velocities
from the NLT and QLT is within 15$\%$. The shaded area indicates the range
of size in which the assumption of constant charge is invalid due to strong
charge fluctuations. Dotted vertical lines present in the WNM and CNM denote
the critical size $a_{\cri}$ corresponding to the cutoff scale $k_{c}$ of turbulence.}
\label{Vcomp}
\end{figure}

Figure \ref{Vcomp} shows grain velocities obtained using the NLT and QLT
for the CNM, WNM and WIM. Both silicate and graphite grains are considered.
Grain velocities obtained using the NLT are generically smaller than those from
the QLT. But the difference is within $15 \%$. The smaller results in the NLT
arise from the fact that, when grain velocities approach $V_{\A}$, a fraction of
turbulence energy is spent to induce transit time acceleration (see also Fig. 5).
The sudden cutoffs (dotted lines) present in the CNM and WNM
correspond to the cutoff scales of turbulence due to collisional and collisionless
damping occurring at the critical size $a_{\cri}$ (see Sec. 4.1).

\begin{figure}
\includegraphics[width=0.45\textwidth]{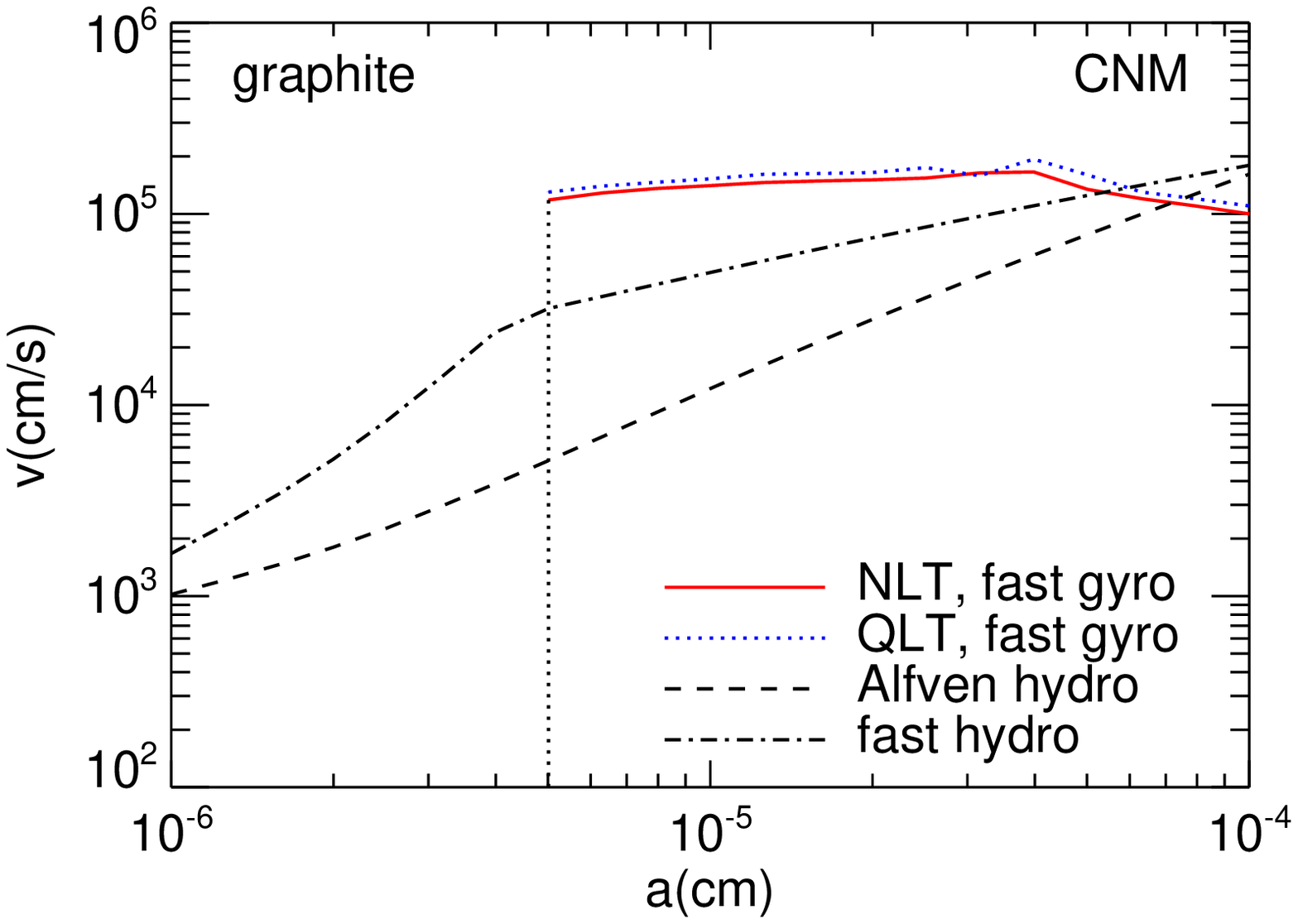}
\includegraphics[width=0.45\textwidth]{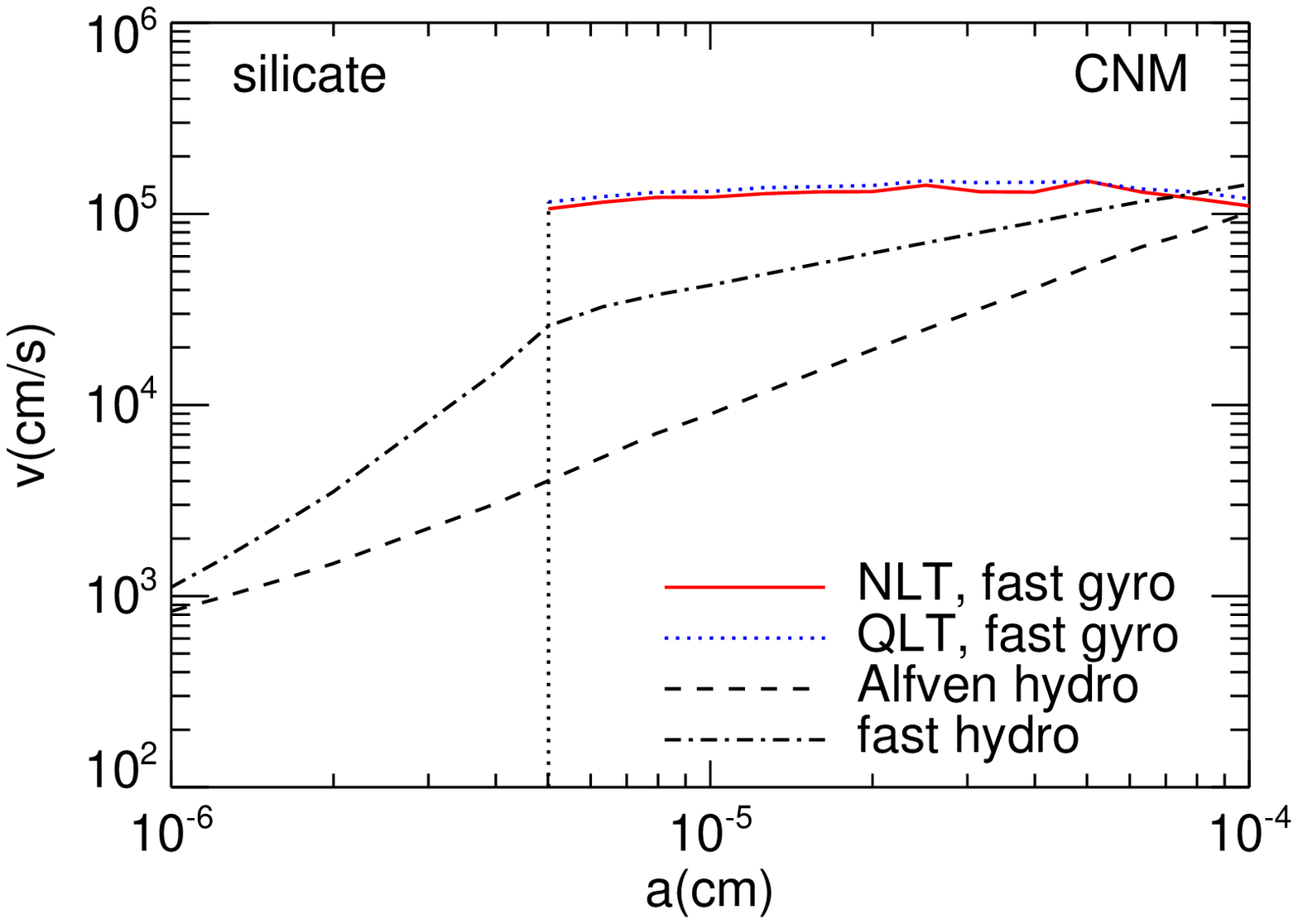}
\caption{Comparison of grain velocities arising from gyroresonant interactions
of fast MHD modes ({\it solid and dotted lines}) with the results arising from
hydrodrag by fast and Alfv\'{e}n modes ({\it dot-dashed and dashed lines}).}
\label{VNQcnm}
\end{figure}

\begin{figure}
\includegraphics[width=0.45\textwidth]{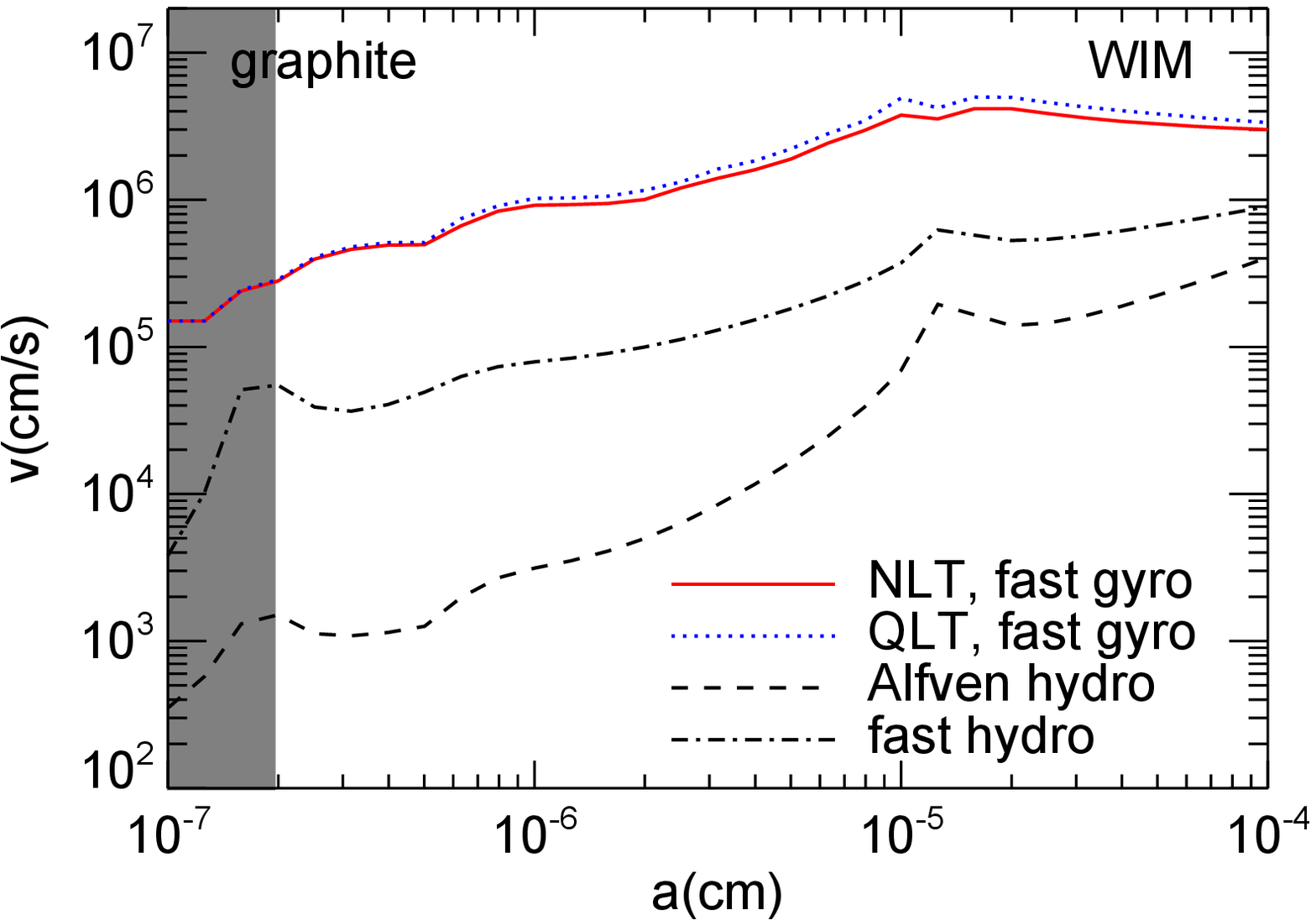}
\includegraphics[width=0.45\textwidth]{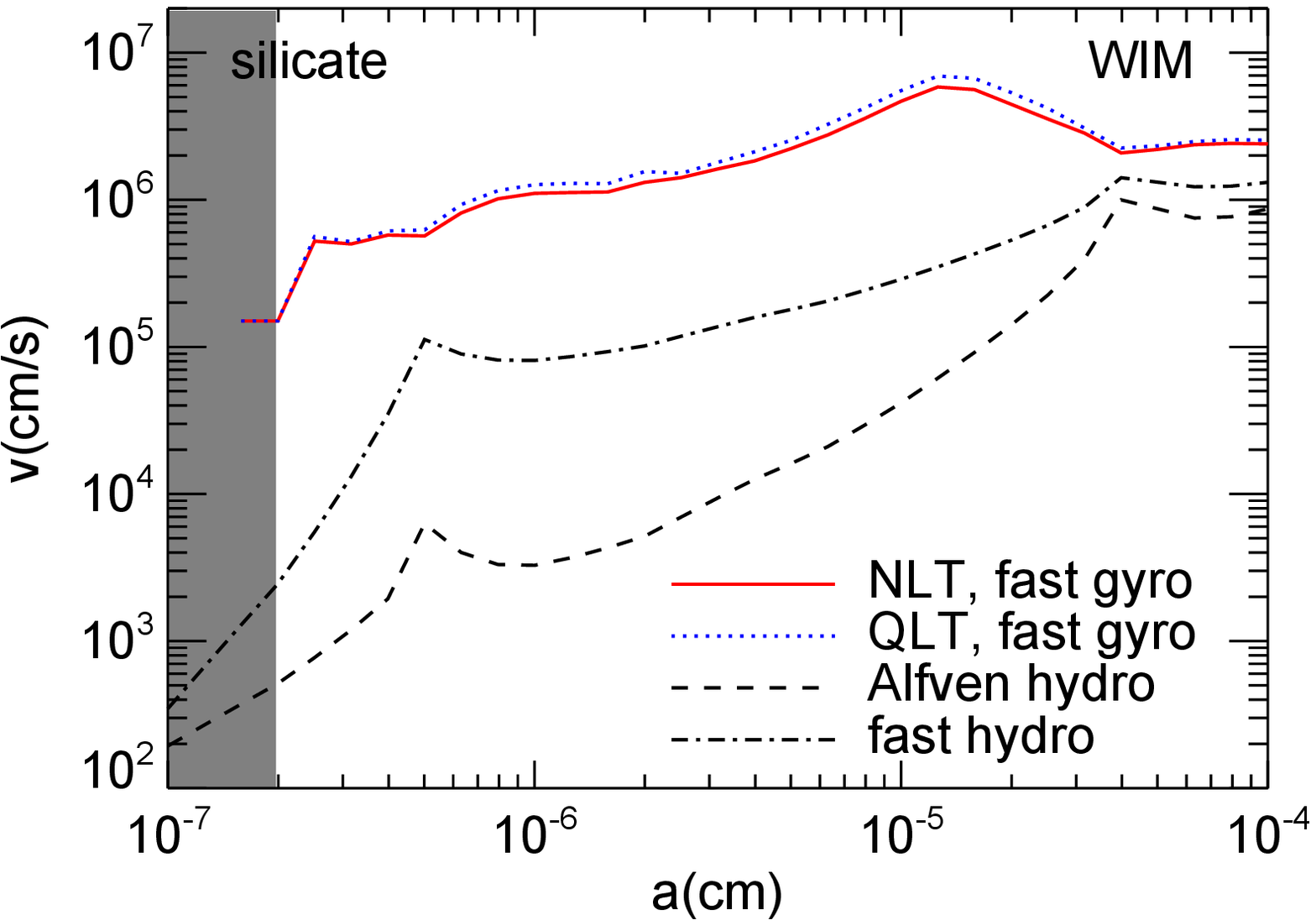}
\caption{Similar to Fig. \ref{VNQcnm} but for the WIM. The shaded area indicates
the range of grain size in which the assumption of constant charge is invalid. Local
maximum velocity for hydro acceleration present in the range $a<10^{-6}$ and
$a>10^{-5}$ cm arise from the change in sign of $\langle Z\rangle$. }
\label{VNQwim}
\end{figure}

Figures \ref{VNQcnm} and \ref{VNQwim} compare grain velocities arising from
fast MHD modes (similar data as Fig. \ref{Vcomp}) with those induced by
Alfv\'{e}n hydro-drag modes (LY02) and fast hydro-drag modes (YLD04)
in the CNM and WIM, respectively.
\footnote{Here only the velocity component perpendicular to magnetic field
is shown.}As expected from earlier studies, gyroresonance acceleration is
dominant for the entire range of grain size in the WIM (Fig. VNQwim).
The rapid variation of grain mean charge \Zmeant present in the WIM
(see Fig. \ref{sigmaZ}) results in the non monotonic
increase of grain velocities from gyroresonance (solid and dotted lines).
Two local maxima in the hydro drag cases correspond to the change in
sign of \Zmeant (see Fig. \ref{sigmaZ}). In the CNM, gyroresonance is
dominant for grain size from $a_{\rm cri}=5\times10^{-6}~\cm$ to
$a\sim 6\times 10^{-5}~\cm$, while acceleration by hydro-drag takes over
for grains smaller than $a_{\cri}$ and larger than $\sim6\times 10^{-5}~\cm$
(see Fig. \ref{VNQcnm}).

\subsection{Acceleration by TTD}
Gyroresonant acceleration tends to drive grain motion in perpendicular direction
to the mean magnetic field (i.e., $\mu=0$). As discussed in Section 3.2, the
broadening of resonance condition in the NLT allows TTD
to operate even at $\mu=0$. In addition to acceleration, the scattering by TTD
can be important, which results in the deviation of $\mu$ from $\mu=0$
(see Yan \& Lazarian 2008). However, it is still uncertain how fast the
pitch angle scattering (both gyroresonance and TTD) by fast MHD modes
is compared to the acceleration. For simplicity, we consider the TTD acceleration
for two limiting cases of {\it efficient} scattering and {\it inefficient} scattering
in which the scattering is more and less efficient than the acceleration, respectively.
In the latter case, the scattering is assumed to be sufficient to alter the adiabatic
invariant of gyromotion.

In the presence of TTD, we take into account the diffusion coefficients
$D_{pp}^{\rm TTD}$ from Equation (\ref{eq:dpt}) for Equation (\ref{eq:dv-dt})
in addition to $D_{pp}^{\rm G}$.

When the pitch angle scattering is efficient, $\mu$ is described by an isotropic
distribution $f(\mu)d\mu=1/2d\mu$. The diffusion coefficient $D_{p}(p,\mu)$
is replaced by its average value over the isotropic distribution. $f(\mu)$.

\begin{figure}
\includegraphics[width=0.45\textwidth]{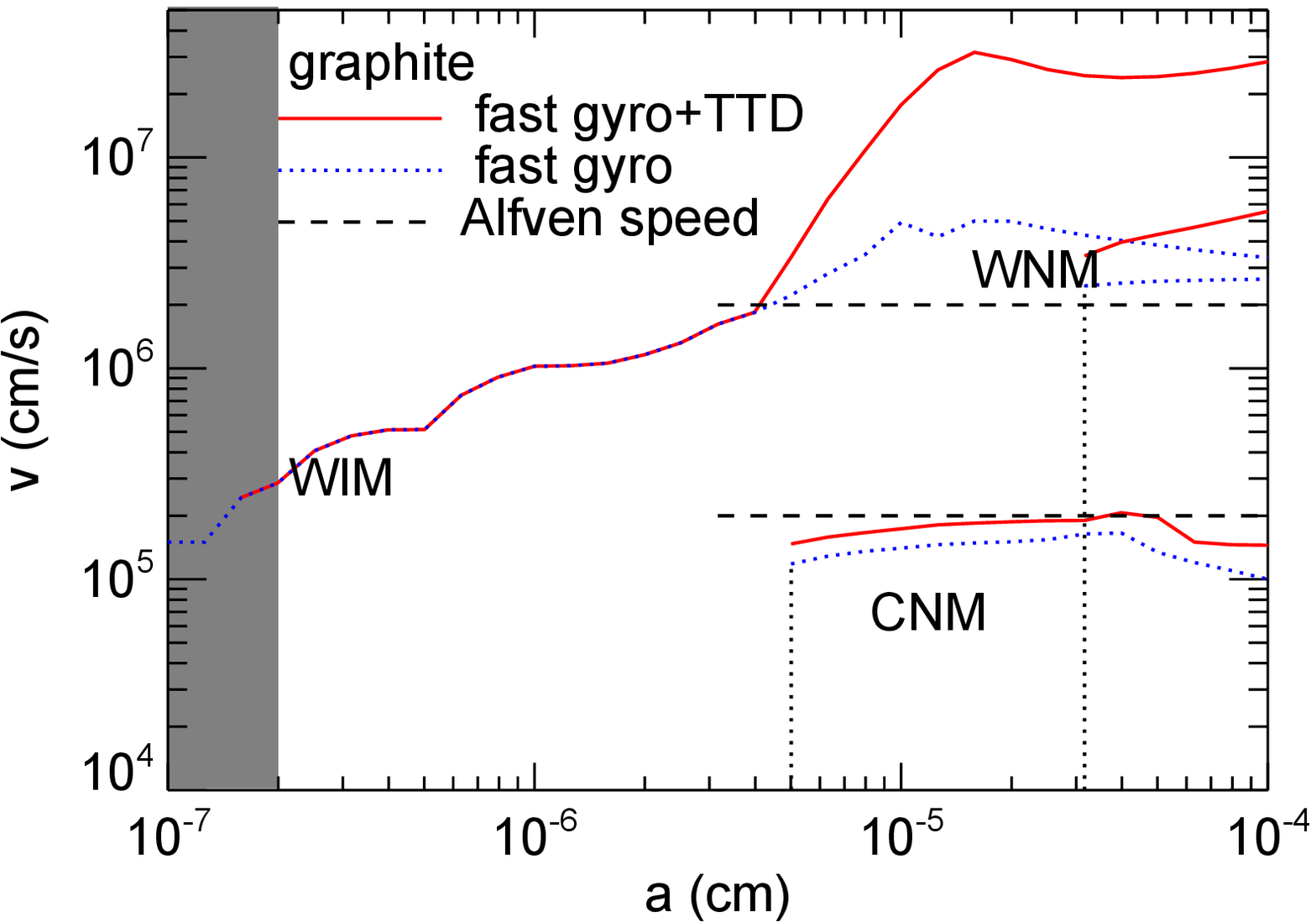}
\includegraphics[width=0.45\textwidth]{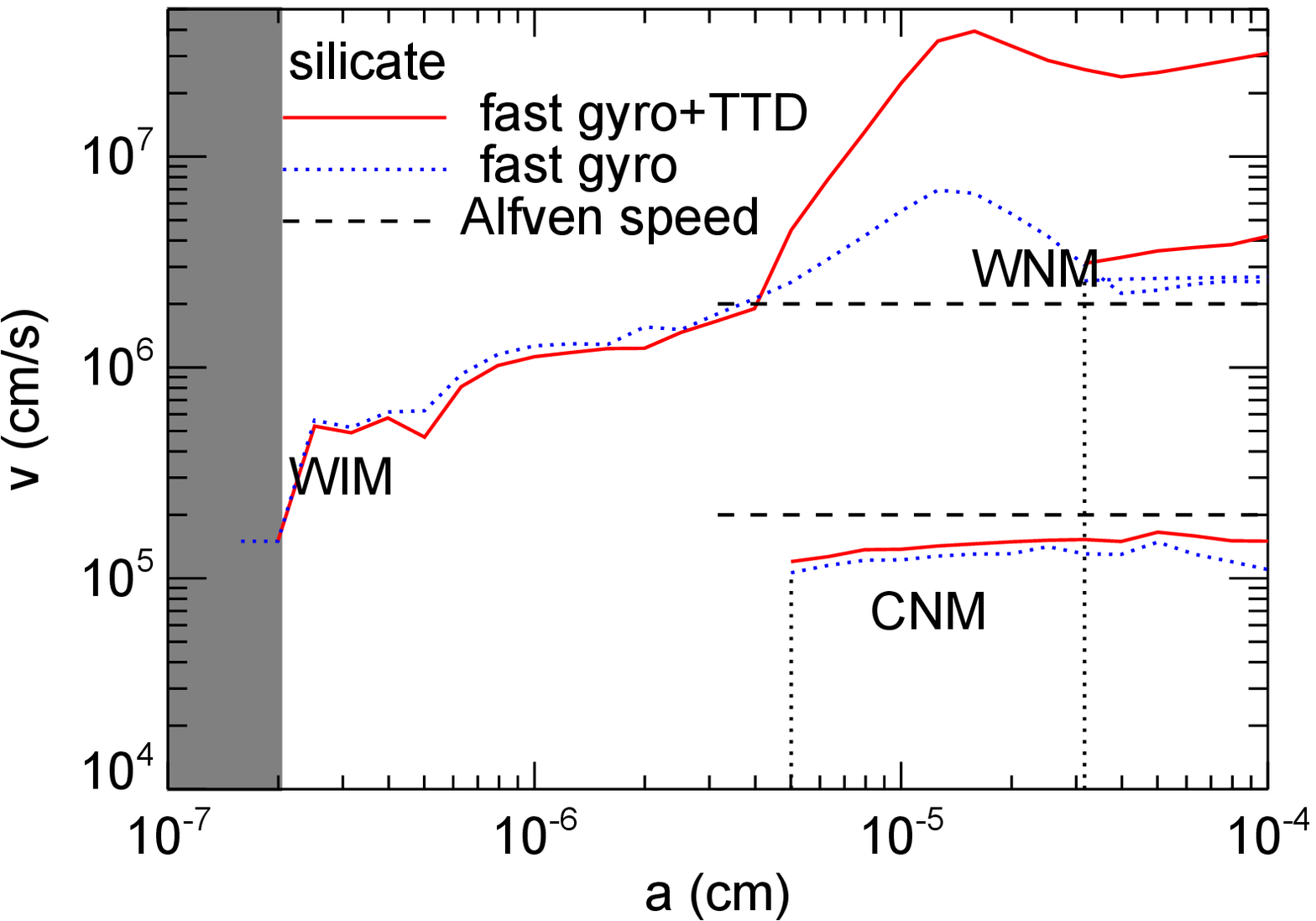}
\caption{ Grain velocities relative to gas arising from gyroresonant acceleration
({\it dotted line}) and gyroresonant acceleration plus transit time acceleration
({\it solid line}), as a function of grain size for different environments.
The case of efficient scattering is considered.}
\label{VGT}
\end{figure}

Figure \ref{VGT} compares grain velocities relative to gas obtained
using the NLT from gyroresonance ({\it dot line}), and
gyroresonance plus TTD by fast MHD modes for silicate grains ({\it upper})
and graphite grains ({\it lower}) assuming the efficient pitch angle scattering.
In the WIM conditions, the TTD acceleration is negligible for grains smaller than
$\sim 4\times 10^{-6}\cm$ for which $v<V_{\A}$. The efficiency of TTD begins
to increase rapidly with $a$ when $v\sim V_{\A}$. TTD can increase grain
velocities to an order of magnitude higher than gyroresonance.
The effect of TTD is less important in the WNM than the WIM, but still
considerable. For the CNM, TTD acceleration is rather marginal
because grains are moving with sub-Alfv\'{e}nic velocities, less than the
threshold for TTD.

In the case of inefficient pitch angle scattering, the cosine of the grain
pitch angle $\mu=0$ is assumed as a result of gyroresonance acceleration.
We found that grain velocities are $\sim 10\%$ larger than the results for
the case of efficient scattering. This seems counterintuitive because
the scattering is required to alter the adiabatic invariant in gyromotion.
However, here we neglected that effect of pitch angle scattering, and
the situation is merely related to the fact that the diffusion coefficient
$D_{pp}$ averaged over $\mu$ is slightly lower than $D_{pp}$ at
$\mu=0$ (see Fig. 3).

\section{Stochastic acceleration by low frequency Alfv\'{e}n waves}
\subsection{General consideration}

The stochastic acceleration by low frequency Alfv\'{e}n waves with
$\omega < \Omega$ at the gyro-scale $k_{\perp}\sim \rho^{-1}$ was studied
in Chandran et al. (2010) for ion heating in the solar wind. Here we consider
the effect of low frequency Alfv\'{e}n waves on dust acceleration in the ISM.
For the low frequency Alfv\'{e}n waves, resonance acceleration is inefficient
because the resonance condition $\omega-k_{\|}v\mu=n\Omega$
is not satisfied. Indeed, in low-$\beta$ plasma, the ion thermal velocity
$v_{T}\ll V_{\A}$, and $\omega=k_{\|} V_{\A}$ for Alfv\'{e}n waves,
we have $\omega-k_{\|}v\mu\ll 0$.

Let $\delta v$ and  $\delta B$ be the rms amplitudes of velocity and magnetic
field at the gyro-scale $k_{\perp}\rho\sim 1$. The electric field induced by
plasma perturbations with velocity $\delta v$ in the direction perpendicular
to the mean magnetic field has the rms amplitude
\bea
\delta E\simeq\frac{\delta v B_{0}}{c},\label{eq:dE}
\ena
and the potential is written as
\bea
\delta \Phi\simeq \rho \delta E.\label{eq:dphi}
\ena
The electric field in the plane perpendicular to $\Bv_{0}$ results in the
acceleration of grains in the perpendicular direction.

Combining these above equations, we obtain
\bea
q\delta\Phi\simeq q\frac{\rho \delta vB_{0}}{c}=mv_{\perp}\delta v,\label{eq:dPhi}
\ena
where $m$ is mass of the charged particle and $\rho=mcv_{\perp}/qB_{0}$
is the gyro radius.

The fractional increase of energy after a single gyro-period is calculated by
\bea
\frac{q\delta \Phi}{mv_{\perp}^{2}/2}\simeq 2\epsilon,
\ena
where 
\bea
\epsilon=\frac{\delta v}{v_{\perp}},\label{eq:epsilon}
\ena

When $\epsilon \ll 1$, also corresponding to $\delta B\ll B_{0}$ due to
the assumption of low $\beta$ plasma, the increase of energy per
gyroperiod is negligible because of adiabatic invariant for the magnetic
moment $mv_{\perp}^{2}/2B_{0}$. As $\epsilon$ increases to unity, the
particle energy changes substantially after one gyroperiod. The guiding
center becomes chaotic when $\epsilon$ exceeds some threshold value,
so that the perpendicular acceleration becomes important.

The increase of energy depends only on the amplitude of perturbation at
the scale of gyroradius. The increase of energy per unit of time  per unit
of mass is defined as
\bea
Q_{\perp}=\frac{v_{\perp}^{2}}{\tau_{\rm acc}},\label{Qperp1}
\ena
where $\tau_{\rm acc}$ is the time for the particle kinetic energy $K_{\perp}$
to increases by a factor of $2$.
We can estimate $\tau_{\rm acc}$ using the diffusion coefficient $D_{K}$
as follows:
\bea
\tau_{\rm acc}=\frac{4K_{\perp}^{2}}{D_{K}}=\frac{m^{2}v_{\perp}^{4}}{D_{K}},\label{taui}
\ena
where
\bea
D_{K}=\frac{\left(\Delta K_{\perp}\right)^{2}}{\Delta t}=m^{2}v_{\perp}^{2}\omega_{\eff}^{2}\delta v\rho,\label{DK}
\ena
and $\omega_{\eff}=\delta v/\rho$ is the effective frequency of gyroscale
fluctuations. Here we have taken $\Delta K_{\perp}\simeq mv_{\perp}\delta
v=mv_{\perp}\omega_{\eff}\rho$ from Equation (\ref{eq:dPhi}),
and $\Delta t=\rho/\delta v$ is the time required for the guiding center to
move by a distance equal to the gyro radius $\rho$.

Taking use of Equations (\ref{taui}) and (\ref{DK}) for (\ref{Qperp1}), we obtain
\bea
Q_{\perp}\simeq \omega_{\eff}^{2}\delta v\rho\simeq\frac{\delta v^{3}}{\rho}.\label{Qperp2}
\ena

When $\epsilon$ is sufficiently small, the variation in the particle energy
is correlated over long time, so the assumption for diffusive approximation
may not be adequate. Thus, the actual energy gain is substantially smaller
than the value obtained by Equation (\ref{Qperp2}). To account for this
reduction, a damping function ${\rm exp}\left(-c_{2}/\epsilon\right)$ is
introduced, and Equation (\ref{Qperp2}) can be rewritten as
\bea
Q_{\perp}=\frac{c_{1}(\delta v)^{3}}{\rho}{\rm exp}\left(-\frac{c_{2}}
{\epsilon}\right),\label{eq:qper}
\ena
where $c_{1}$ and $c_{2}$ are dimensionless constants that depend on the
nature of fluctuations. Below we assume $c_{1}=0.75$ and $c_{2}=0.34$ for
the fluctuations by low frequency Alfv\'{e}n waves as in Chandran et al. (2010).

\subsection{Grain velocities for the ISM}

Let assume that the Alfv\'{e}nic turbulence in the ISM follows the scaling
\bea
\delta v_{\perp}=\alpha V_{\A}\left(\frac{\l_{\perp}}{L}\right)^{a}\label{eq:deltav}
\ena
where $l_{\perp}\sim \rho_{\d}$ and $L$ are gyro scale and the injection scale.
$\alpha$ and $a=(c_{3}-1)/2$ with $c_{3}$ being the slope of turbulence
power spectrum are dimensionless, derived from the properties of turbulence.
For sub-Alfv\'{e}nic turbulence,  $\alpha<1$. Using Equation (\ref{eq:epsilon})
for dust grains, i.e., $\epsilon_{\d}=\delta v_{\perp}/v_{\perp,d}$, we obtain
\bea
\epsilon_{d}=\alpha \left(\frac{B^{2}}{8\pi n_{\rm H}k_{\rm B}T_{\perp}} \right)^{(1-a)/2}\frac{A^{(1+a)/2}}{Z^{a}}\left(\frac{d_{p}}{L}\right)^{a},\label{eq:epschao}
\ena
where $p$ denotes proton, and $d$ denotes dust, $A=m/m_{p}$,
$d_{p}=V_{\A}/\Omega_{p}$, and the perpendicular temperature is defined as
$kT_{\perp}=m v_{\perp,\d}^{2}/2$.

Using Equations (\ref{eq:deltav}) and (\ref{eq:epschao}) for Equation (\ref{eq:qper}),
we obtain the rate of energy gain per a grain of mass $m$
\bea
A(v_{\perp})=m\frac{c_{1}V_{\A}^{3}}{L}{\rm exp}\left(-\frac{c_{2}}
{\epsilon_{d}}\right).\label{eq:qper1}
\ena
For calculations, we assume $c_{3}=5/3$ and $a=1/3$ for Alfv\'{e}nic
turbulence above the gyro scale.

\begin{figure}
\includegraphics[width=0.45\textwidth]{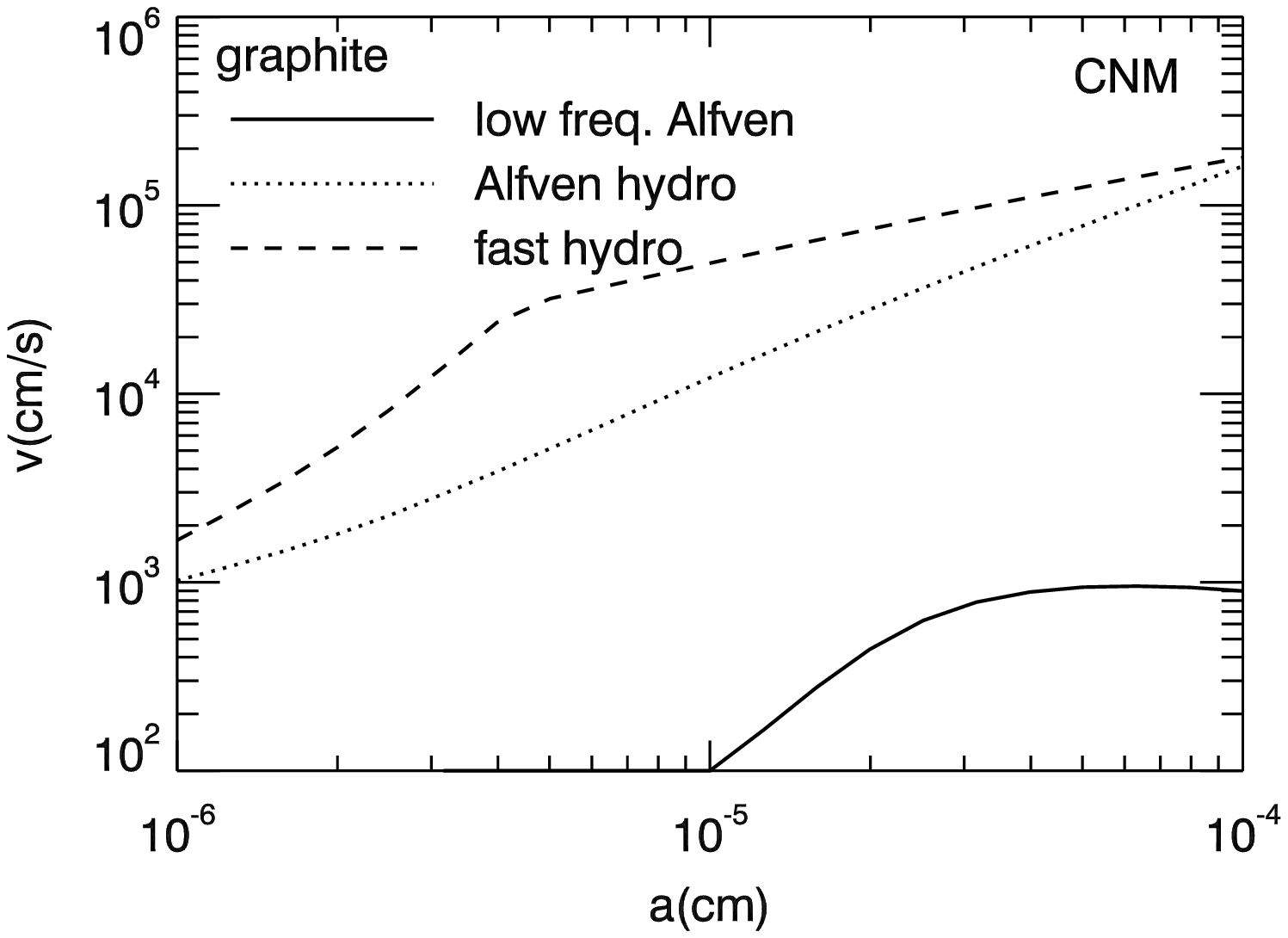}
\includegraphics[width=0.45\textwidth]{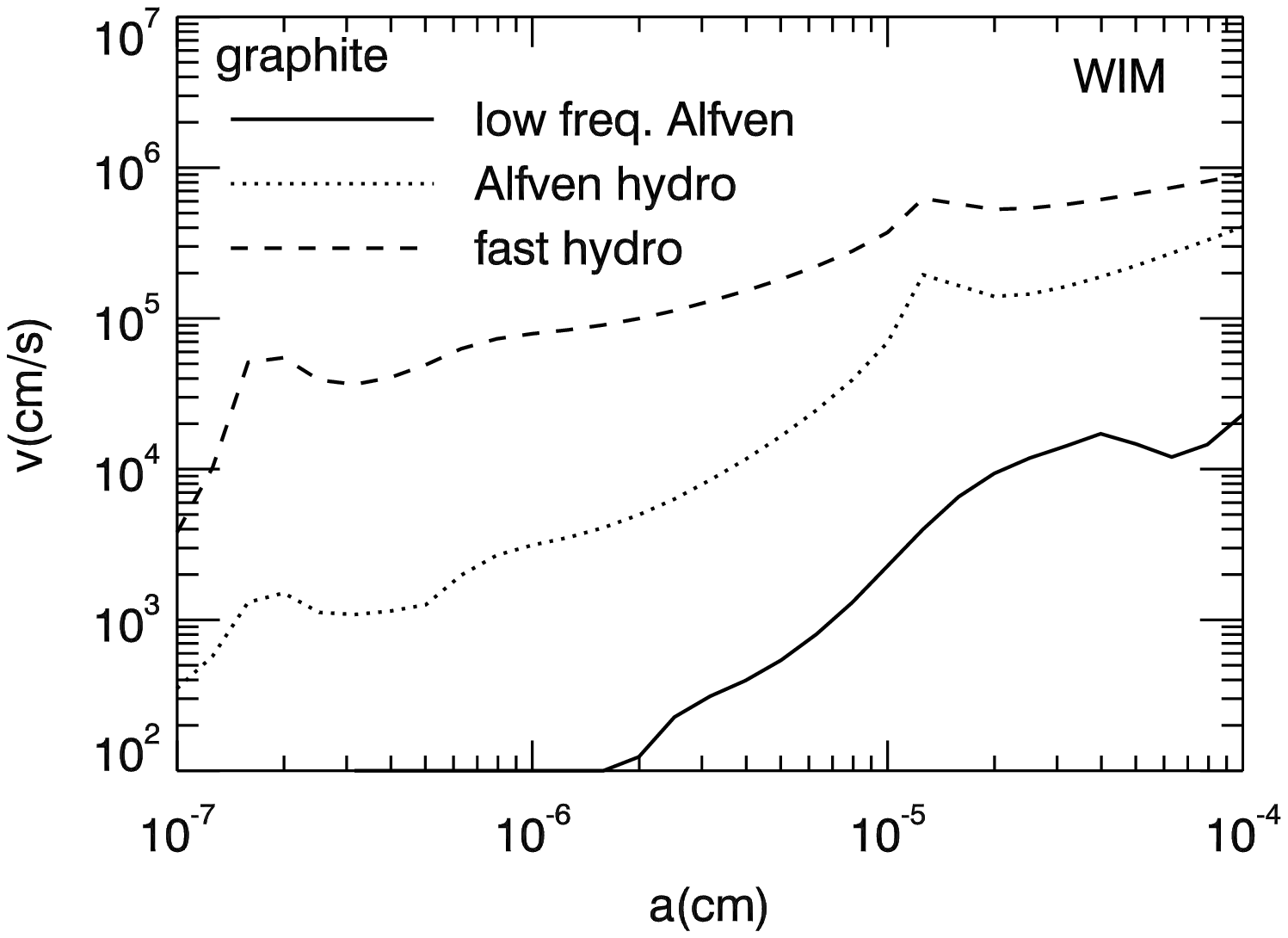}
\caption{Grain velocity due to stochastic acceleration by low frequency
Alfv\'{e}n waves (solid lines) compared to the acceleration by hydrodynamic
drag from Alfv\'{e}n modes (dotted lines) and fast modes (dashed lines),
for the CNM ({\it upper}) and WIM ({\it lower}). The stochastic acceleration
is much less efficient than the latter.}
\label{f4}
\end{figure}

Using the parameters for the ISM in Table 1, we calculate the grain velocity
arising from the chaotic acceleration by low frequency Alfv\'{e}n waves in
Figure \ref{f4} for the CNM and WIM. We show that the chaotic acceleration
by low frequency Alfv\'{e}n waves is subdominant to the fast and Alfv\'{e}nic
hydrodynamic drag. Obviously, it is much less important than gyroresonance
and TTD by fast modes. The possible reason is that the low frequency
Alfv\'{e}n waves cascade faster to small scale than the fast modes.

\section{Discussion}

\subsection{Related works on dust grain acceleration}

The acceleration of dust grains by incompressible MHD turbulence was first studied
by Lazarian \& Yan (2002). Yan \& Lazarian (2003) studied grain acceleration
in compressible MHD turbulence, and discovered a new acceleration mechanism
based on gyroresonant interactions of grains with waves. This acceleration
mechanism increases grain velocities in perpendicular direction to the mean
magnetic field. YLD04 computed grain velocities arising from gyroresonance
by fast MHD modes using quasi-linear theory (QLT), and compared the obtained
results with different mechanisms, for various ISM phases. They found that the
gyroresonance is the most efficient mechanism for grain acceleration in the ISM.

The effect of large scale compression on grain acceleration is shown
by Yan (2009) to be less important than the gyroresonance in the ISM
conditions, unless the grains move with super-Alfv\'{e}nic velocities.

For very small grains (e.g., polycyclic aromatic hydrocarbons and nanoparticles),
Ivlev et al. (2010) sketched a new mechanism of grain acceleration due to
electrostatic interactions of grains with fluctuating charge and provided
rough estimates of grain velocities in the ISM. Hoang \& Lazarian (2011)
quantified this mechanism using Monte Carlo simulations of charge
fluctuations. They found that charge fluctuations can accelerate grains to
several times their thermal velocities.

\subsection{NLT for gyroresonance acceleration}

We have revisited the treatment of gyroresonance acceleration for charged
grains due to MHD turbulence by accounting for the fluctuations of grain guiding
center from a regular trajectory along the mean magnetic field (i.e.
NLT limit). The fluctuations of the guiding center result in the broadening of
resonance conditions-- a Delta function is replaced by a Gaussian function.
Such broadening of resonance condition allows some fraction of wave energy
spent through the TTD acceleration. As a result, grain velocities due to
gyroresonance acceleration are in general decreased by $\sim 15\%$
in the NLT limit.

\subsection{Transit time damping acceleration}

TTD acceleration is believed to be important when the parallel component of
grain velocity along the magnetic field exceeds the Alfv\'{e}n speed $V_{\A}$.
Although gyroresonance acceleration by fast modes can accelerate grains to
$v\ge V_{\A}$, their resulting velocity mostly perpendicular to the magnetic field,
i.e. $\mu=0$, makes TTD unfavored because the resonance condition
$\delta (\omega-k_{\|}v_{\|})$ is not satisfied. Indeed, we found that TTD is
efficient for $\mu>V_{\A}/v$ and negligible for $\mu<V_{\A}/v$ in the QLT limit.
This feature is consistent with the result for acceleration of cosmic rays in
Schlickeiser \& Miller (1998).

The situation changes when the fluctuations of the guiding center are taken
into account in the NLT. For this case, the resonance condition is broadened
beyond the $\delta$ function, and can be described by a Gaussian function.
As a result, TTD acceleration becomes important for $\mu <V_{\A}/v$,
including $90^{\circ}$ pitch angle.

In addition to acceleration, TTD also induces the grain pitch angle scattering,
which is dominant over the scattering by gyroresonance. Since the efficiency
of the TTD scattering is uncertain, we considered in the paper two limiting
cases of inefficient and efficient scattering in which the scattering is less and
more efficient than the acceleration. The pitch angle is equal to $90^{\circ}$
in the former, and isotropic in the latter.

When the scattering is more efficient than the acceleration, we showed that
for the WNM and WIM, the TTD acceleration can increase substantially the
grain velocity compared to results arising from gyroresonance.
Particularly, for grains larger than $5\times 10^{-6}$ cm in the WIM, TTD
acceleration is an order of magnitude greater than the gyroresonance
acceleration. TTD is clearly more efficient than the betatron acceleration studied
in Yan (2009). In the CNM, TTD acceleration is limited because the gyroresonance
acceleration is not able to speed up grains to super Alfv\'{e}nic stage.
When the scattering is less efficient than the acceleration, grain velocities are
within $10\%$ lower than the results for the efficient scattering case.

\subsection{Stochastic acceleration}
We study also the effect of low frequency  ($\omega< \Omega$) Alfv\'{e}n
waves on dust grains in the ISM. We show that the stochastic acceleration
by low frequency Alfv\'{e}n waves is subdominant to the gyro-resonance
acceleration and TTD acceleration by fast modes. This may arise from the
fact that low frequency Alfv\'{e}n waves cascade faster to small scale than
the fast modes in MHD turbulence.

\subsection{Implication to dust coagulation and shattering and alignment}

Hirashita \& Yan (2009) adopted grain velocity due to gyroresonance
acceleration from YLD04 to model grain size distribution in the different ISM
conditions. Hirashita et al. (2010) studied grain coagulation and shattering in
the WIM for large dust grains ejected from Type II supernova. They showed
that the shattering of large dust grains due to turbulence plays an important
role in producing small size population that modifies extinction curves in
starburst galaxies.

The threshold velocity for the grain shattering is a function of the grain size:
\bea
v_{\rm shat}=2.7\left(\frac{a}{10^{-7}~\cm}\right)^{-5/6} {\rm km}~\s^{-1},\label{v_cri}
\ena
where $v_{\rm dd}$ is the relative velocity of dust grains (Chokshi et al. 1993).
If $v_{\rm dd}< v_{\rm shat}$, the grains collide and stick together. When
$v_{\rm dd} > v_{\rm shat}$, the collisions with high velocity produce shock
waves inside the grains, and shatter them in smaller fragments. For
$v_{\rm dd}\geq 20$ km/s, the evaporation of the dust grain occurs and the
grains are destroyed.

When TTD is accounted for, grains larger than $5\times 10^{-6}$ in the WIM
and WNM may undergo efficient shattering because $v_{\rm dd}>V_{\A}=20$
km $\s^{-1}$.

The effects of high velocities we obtain on dust grain alignment require further
studies. Recent research has shown that the classical grain alignment theory
of Davis \& Greenstein (1951) (see also Lazarian 1995; Roberge \& Lazarian
1999 for more recent quantitative studies of the process) is subdominant to the
radiative torque (RAT) model (see Dolginov \& Mitrofanov 1976; Draine \&
Weingartner 1996; Lazarian \& Hoang 2007a). This model, however, was
criticized in Jordan \& Weingartner (2009) who appealed to the results of
gyroresonance acceleration of grains in  Yan \& Lazarian (2003) and claimed
this means that fast moving grains will be randomized as their charge fluctuates.
Our results indicate that grains can be accelerated to even faster velocities,
which could make the problem for the alignment more severe. However, we
believe that the claim about suppression of the RAT alignment for fast moving
grains is a result of the confusion on the nature of the RAT alignment.
We plan to address this issue elsewhere. At the same time
high velocities of grains may induce another mechanical alignment of irregular
grains as it described in Lazarian \& Hoang (2007b). This alignment does
require further studies.

\section{Summary}

In the present paper, we study the resonance acceleration of charged grains
by fast modes and stochastic acceleration by Alfv\'{e}n waves in MHD turbulence.
Our main results are summarized as follows.

1. We revisit the treatment of gyroresonance acceleration of charged grains
in compressible MHD turbulence by taking into account the fluctuations of
grain guiding center from the regular trajectory along the mean field.
Gyroresonance interactions by fast modes can accelerate large grains to
super-Alfv\'{e}nic speed. We found that grain velocities are lower
by $15\%$ in the NLT than the QLT.

2. We investigate the effect of transit time damping (TTD) by fast modes
for super-Alfv\'{e}nic grains. We found that the fluctuations of grain guiding
center allow TTD to occur not only within the range of the cosine of the grain
pitch angle $\mu>V_{\A}/v$ as expected by the QLT, but also for $\mu<V_{\A}/v$.
We show that the TTD acceleration can increase grain velocities by an order of
magnitude compared to the results arising from gyroresonance mechanism
Thus, TTD is the most efficient acceleration mechanism for super-Alfv\'{e}nic grains.

3. The stochastic acceleration due to low frequency Alfv\'{e}n waves is
inefficient for dust grains in the ISM conditions.

\acknowledgements
AL thanks Alexander von Humboldt Foundation. R.S. acknowledges the
support  from the Deutsche Forschungsgemeinschaft through grants Schl
201/19-1 and Schl 201/23-1. TH and AL acknowledge the support of the Center for
Magnetic Self-Organization. We thank Bruce Draine for providing us the data
of grain charge distribution and Huirong Yan for valuable comments.
We thank the anonymous referee for her/his useful comments that improve the paper.
\appendix

\section{A. Turbulence cascade and damping}

We summarize here major damping processes for the MHD turbulence. First,
we begin with the cascade of different MHD turbulence modes.
 
\subsection{A.1. Turbulence cascade} 
Turbulence cascades from large scale to small scale. The cascade rate
depends on the scale of the eddy and on the type of turbulence mode.
For Alfv\'{e}n and pseudo-Alfv\'{e}n (slow) modes, the cascade time at the scale $k$ is equal to
the eddy turn-over time:
\bea
\tau_{\cas}^{A,s}=\frac{1}{k_{\|} V_{\A}}\equiv \frac{1}{k_{\perp}v_{k}}.\label{tcas_Alfven}
\ena
where the critical balance condition  (GS95) for the cascade 
parallel and perpendicular to the mean magnetic field 
\bea
	k_{\|}v_{\A}=k_{\perp}v_{\perp}
\ena
has been used. Here $v_{\perp}\sim v_{k}$ because the Alfv\'{e}n mode has
perturbation velocity $v_{k}$ perpendicular to the mean magnetic field.

Fast modes cascade a bit slower than the eddy turn over,
and the cascade time is given by
\bea
\tau_{\cas}^{f}=\left(\frac{l}{v_{k}}\right)\left(\frac{v_{ph}}{v_{k}}\right)=\frac{k^{-1}V_{f}}{v_{k}^{2}}\sim \left(\frac{k}{L}\right)^{1/2}\frac{\delta V^{2}}{v_{ph}},\label{tcas_fast}
\ena
where $v_{ph}$ is the phase speed. In low-$\beta$ plasma, $v_{ph}\equiv v_{\A}$
and $v_{ph}\equiv c_{s}$ in high-$\beta$ plasma. 

\section{B. Turbulence damping}
\subsection{B.1. Collisional damping}
\subsubsection{B.1.1. Neutral Ion collision damping}
In partially ionized gas, the MHD turbulence energy is dissipated through
ion-neutral collisions. Let $\lambda_{n}$ be the mean free path  of a
neutral. For scales $l>\lambda_{n}$, the ion-neutral collisions can damp
the turbulence at a damping rate
\bea
\Gamma_{ni}\sim {\nu_{n}l^{-2}}\sim \frac{n_{n}}{n_{\gas}}(\lambda_{n}v_{n})k^{2},\label{gam_ni}
\ena
where $l\sim k^{-1}$ is the scale of interest, $\nu_{n}$ is the effective viscosity due to neutrals,
 and $n_{\gas}=n_{i}+n_{n}$ is the gas number density.

\subsubsection{B.1.2. Viscous damping}
In fully ionized gas, the turbulence can get damped due to viscous damping or 
collisionless damping depending on whether regimes are collisional or collisionless. 
The Coulomb mean free path of the thermal plasma is given by 
\bea
\lambda_{\rm Coul}=v\tau_{\rm Coul}=\frac{v}{n v\sigma}=\frac{m^{2}v^{4}}{n\pi e^{4}}=9\times 10^{7}\left(\frac{T_{\gas}}{10^{7}~\K}\right)^{2}\left(\frac{10^{10}\cm^{-3}}{n}\right)~\cm,
\ena
where $\sigma=\pi r_{\min}^{2}=\pi \left(e^{2}/mv^{2}\right)$ has been used.

The viscous damping is important for scales between the injection scale $L$ and 
the mean free path $\lambda_{\rm Coul}$. For scales $l<\lambda_{\rm Coul}$, the collisionless
damping is dominant.

\subsection{B.2. Collisionless regime}
\subsubsection{B.2.1. ion viscosity damping}
The motion in the perpendicular direction to the magnetic field is suppressed,
so the viscosity is much smaller in the perpendicular direction compared to the
viscosity in the parallel direction, i.e. $\eta_{\perp}\sim \eta_{0}/(\Omega_{i}\tau_{\rm Coul})^{2}$
where $\tau_{\rm Coul}\sim \lambda_{\rm Coul}/v_{th}$ is the Coulomb collision time for ion
and $v_{th}$ is the ion thermal velocity, and $\eta_{0}=0.96 nk_{B}T\tau_{\rm Coul}$ is the longitudinal viscosity for motion  
along the magnetic field (see  Braginskii 1965). Petrosian et al. (2006)  derived
\bea
\Gamma_{ivisc}=k_{\perp}^{2}\frac{\eta_{0}}{6n_{i}m_{i}},\label{gamma_ivis}
\ena
for $\beta\ll 1$ and 
\bea
\Gamma_{ivisc}=\frac{k^{2}\eta_{0}(1-3\cos\theta^{2})^{2}}{6n_{i}m_{i}}
\ena
for $\beta \gg 1$.
For fast modes, the cut-off due to viscous damping is obtained in YL08:
\bea
k_{c}L=x_{c}(1-\cos\beta^{2})^{-2/3}
\ena
for $\beta\ll 1$ and 
\bea
k_{c}L=x_{c}(1-3\cos\beta^{2})^{-4/3}
\ena
for the high $\beta$ plasma, where $x_{c}=\left(6\rho\delta V^{2}L/(\eta_{0}V_{\A})\right)^{2/3}$.

For fast modes, the cut-off scale of turbulence is given by 
\bea
k_{c}L=\left(\frac{L\beta}{18}\right)^{-1/3}\left(\frac{\lambda_{Coul}\sin\theta^{2}}{M_{\A}^{2}}\right)^{-2/3}=\frac{4}{\beta^{1/3}}\left(\frac{M_{\A}}{\sin\theta}\right)^{4/3}\left(\frac{L}{10^{8}\cm}\right)^{2/3}\left(\frac{10^{10}\cm}{n}\right)\left(\frac{10^{7}\K}{T_{\gas}}\right).\label{kc_ivisc}
\ena

\subsubsection{B.2.2. Landau damping}
In collisionless plasma, the turbulence is damped due to the Landau damping.
The damping rate due to the Landau damping is given by (see YL08)
\bea
\Gamma_{ncol}=\frac{\sqrt{\pi\beta}\sin\theta^{2}}{2\cos\theta}kv_{\A}\left[\left(\frac{m_{e}}{m_{i}}\right)^{1/2}\exp-\left(\frac{m_{e}}{\beta m_{i}\cos\theta^{2}}\right)+5\exp-\left(\frac{1}{\beta\cos\theta^{2}}\right)\right].\label{gamma_ncol}
\ena
The cut-off scale of the turbulence is obtained by equating the damping rate to the 
cascading rate. From Equations (\ref{tcas_fast}) and (\ref{gamma_ncol}) we
obtain
\bea
k_{c}L=\frac{4M_{\A}^{2}m_{i}\cos\theta^{2}}{\pi m_{e}\beta\sin\theta^{4}}\exp\left(-\frac{2m_{e}}{\beta m_{i}\cos\theta^{2}}\right)
\ena

\section{C. NLT of gyroresonance acceleration of dust grains}
Below we describe the NLT for resonance acceleration of dust grains in MHD turbulence.

\subsection{C.1. Fokker-Plank coefficients}
The QLT assumes that the guiding center of the charged particles is regular motion along the 
uniform magnetic
 field and that the gyro-orbit is not perturbed. 

The Fokker-Planck diffusion coefficients (Jokipii 1966; Schlickeiser \& Miller 1998) are given as
\begin{eqnarray}
\left(\begin{array}{c}
D_{\mu\mu}\\
D_{pp}\end{array}\right) & = & {\frac{\pi\Omega^{2}(1-\mu^{2})\delta V^{2}}{u_{B}}}\int_{\bf k_{min}}^{\bf k_{max}}dk^3R_{n}\left(k_{\|}v\mu-\omega+ n\Omega\right)\left(\begin{array}{c}
\left(1+\frac{\mu V_{ph}}{v\zeta}\right)^{2}\\
m^{2}V_{A}^{2}
\end{array}\right) \label{eq:gyroapp}\\
 &  & \left\{ (J_{2}^{2}({\frac{k_{\perp}v_{\perp}}{\Omega}})+J_{0}^{2}({\frac{k_{\perp}v_{\perp}}{\Omega}}))\left[\begin{array}{c}
M_{{\mathcal{RR}}}({\mathbf{k}})+M_{{\mathcal{LL}}}({\mathbf{k}})\\
K_{{\mathcal{RR}}}({\mathbf{k}})+K_{{\mathcal{LL}}}({\mathbf{k}})
\end{array}\right]\right.\nonumber \\
 & - & 2J_{2}({\frac{k_{\perp}v_{\perp}}{\Omega}})J_{0}({\frac{k_{\perp}v_{\perp}}{\Omega}})\left.\left[e^{i2\phi}\left[\begin{array}{c}
M_{{\mathcal{RL}}}({\mathbf{k}})\\
K_{{\mathcal{RL}}}({\mathbf{k}})
\end{array}\right]+e^{-i2\phi}\left[\begin{array}{c}
M_{{\mathcal{LR}}}({\mathbf{k}})\\
K_{{\mathcal{LR}}}({\mathbf{k}})
\end{array}\right]\right]\right\}, \nonumber 
\end{eqnarray}
where $u_{B}=B_{0}^{2}/8\pi$, $|{\bf k_{min}}|=k_{\min}=L^{-1}$, $|{\bf k_{max}}|=k_{\max}$ corresponds 
to the dissipation scale, $\mathcal{R,L}$ refer to the left- and right-circularly
 polarized modes, and $\phi=\tan^{-1}k_x/k_y$. Above $R_{n}$ is the function for
 resonance condition, $V_{ph}$ is the phase speed and $\Omega$ is Larmor frequency. 

The correlation tensors are defined as
\begin{equation}
\begin{array}{c}
\langle B_{\alpha}(\mathbf{k},t)B_{\beta}^{*}(\mathbf{k'},t+\tau)\rangle/B_{0}^{2}
=\delta(\mathbf{k}-\mathbf{k'})M_{\alpha\beta}(\mathbf{k})e^{-\tau/\tau_{k}}\\
\langle v_{\alpha}(\mathbf{k},t)v_{\beta}^{*}(\mathbf{k'},t+\tau)\rangle/V_{A}^{2}
=\delta(\mathbf{k}-\mathbf{k'})K_{\alpha\beta}(\mathbf{k})e^{-\tau/\tau_{k}},
\end{array}
\end{equation}
 where $B_{\alpha,\beta}$, $v_{\alpha,\beta}$ are respectively the
magnetic and velocity perturbation associated with the turbulence, $\tau_{k}$
is the nonlinear decorrelation time and essentially the cascading
time of the turbulence. For the balanced cascade we consider (see
discussion of our imbalanced cascade in CLV02), i.e., equal intensity
of forward and backward waves, $C_{ij}(\mathbf{k})=0$.

The magnetic correlation tensor for Alfv\'{e}nic turbulence is (CLV02),
\begin{eqnarray}
\left[\begin{array}{c}
M_{ij}({\mathbf{k}})\\
K_{ij}({\mathbf{k}})
\end{array}\right]&=&\frac{L^{-1/3}}{12\pi}I_{ij}k_{\perp}^{-10/3}\exp(-L^{1/3}|k_{\parallel}|/k_{\perp}^{2/3}),\nonumber\\
\tau_{k}&=&(L/V_{A})(k_{\perp}L)^{-2/3}\sim \left(k_{\parallel}V_{A}\right)^{-1}\label{anisotropic}\end{eqnarray}
where $I_{ij}=\{\delta_{ij}-k_{i}k_{j}/k^{2}\}$ is a 2D tensor in
$x-y$ plane which is perpendicular to the magnetic field, $L$ is
the injection scale, $V$ is the velocity at the injection scale.
Slow modes are passive and similar to Alfv\'{e}n modes. The normalization
constant is obtained by assuming equipartition $\epsilon_{k}=\int dk^{3}\sum_{i=1}^{3}M_{ii}B_{0}^{2}/8\pi\sim B_{0}^{2}/8\pi$.
The normalization for the following tensors below are obtained in
the same way.

According to CL02, fast modes are isotropic and have one dimensional
energy spectrum $E(k)\propto k^{-3/2}$. In low $\beta$ medium, the
corresponding correlation is (YL03)

\begin{eqnarray}
\left[\begin{array}{c}
M_{ij}({\mathbf{k}})\\
K_{ij}({\mathbf{k}})
\end{array}\right]={\frac{L^{-1/2}}{8\pi}}H_{ij}k^{-7/2}\left[\begin{array}{c}
\cos^{2}\theta\\
1\end{array}\right],~~
\tau_{k}=(k/L)^{-1/2}\times V_{A}/V^{2},
\label{fast}
\end{eqnarray}
where $\theta$ is the angle between $\mathbf{k}$ and $\mathbf{B}$,
$H_{ij}=k_{i}k_{j}/k_{\perp}^{2}$ is also a 2D tensor in $x-y$ plane.
The factor $\cos^{2}\theta$ represents the projection as magnetic
perturbation is perpendicular to $\mathbf{k}$. This tensor is different
from that in Schlickeiser \& Miller (1998). For isotropic turbulence,
the tensor of the form $\propto E_{k}(\delta_{ij}-k_{i}k_{j}/k^{2})$
was obtained to satisfy the divergence free condition $\mathbf{k}\cdot\delta\mathbf{B}=0$
(see Schlickeiser 2002). Nevertheless, the fact that $\delta\mathbf{B}$
in fast modes is in the $\mathbf{k}$-$\mathbf{B}$ plane places another
constraint on the tensor so that the term $\delta_{ij}$ doesn't exist. 

\subsection{C.2. Diffusion coefficients in NLT}

The motion of a charged particle in a magnetic field $\Bv$ consists of the motion of the guiding 
center with respect to the magnetic field $\Bv$ and the motion of the particle about the 
guiding center. In the QLT limit, the guiding center is assumed to
follow regular trajectory with constant pitch angle $\mu$.
In MHD turbulence, $\Bv$ varies with respect to space and time, so $\mu$ changes,
 and $v_{\|}$ and $v_{\perp}$ change accordingly.

In the NLT limit, the dispersion of the pitch angle due to magnetic field fluctuations reads
\bea
\Delta \mu=\frac{\Delta v_{\|}}{v},\label{eq:dmu}
\ena
and
\bea
\frac{\Delta v_{\|}}{v_{\perp}}=\frac{\langle (B-B_{0})^{2}\rangle^{1/4}}
{B_{0}^{1/2}}=\left[\frac{\langle(\delta B_{\|})^{2}\rangle}{B_{0}^{2}}+
O\frac{\langle(\delta B_{\perp})^{2}\rangle}{B_{0}^{2}}\right]^{1/4},~~~\label{eq:Deltav}
\ena
where $B_{0}$ is the mean magnetic field (V\"{o}lk 1975). The dispersion of parallel velocity $\delta v_{\|}$ is 
mainly induced by the fluctuations of the parallel magnetic field $\delta B_{\|}$, while
$\delta B_{\perp}$ is only second order effect.

Since $\mu$ is constant in the QLT, the resonance condition is given by 
$\delta(k_{\|}\mu v-\omega+n\Omega)$. In the NLT, due to the fluctuations of $\mu$ (Eq. \ref{eq:dmu}),
the resonance condition is broadened and described by 
\bea
R_{n}\left(k_{\|}v\mu-\omega+ n\Omega\right)=\frac{\sqrt{\pi}}{k_{\|}\Delta v_{\|}}{\rm exp}\left[-\frac{(k_{\|}v\mu-\omega+ n\Omega)^{2}}{k_{\|}^{2}(\Delta v_{\|})^{2}}\right],~~~\label{eq:Rn}
\ena
where $n=0$ and $\pm 1$ (see YL08; YLP08).

Using Equations (\ref{fast}) for (\ref{eq:gyroapp}) combined with (\ref{eq:Rn}), we obtain
\bea
D_{pp}(\mu)^{\rm G}=\frac{v\sqrt{\pi}\Omega^{2}(1-\mu^{2})m^{2}V_{\A}^{2}M_{\A}^{2}}
{4LR^{2}}\int_{1}^{k_{c}L}x^{-5/2}dx\int_{0}^{1} \frac{d\eta}{\eta\Delta \mu}[J_{0}^{2}(w)+J_{2}^{2}(w)]
{\rm exp}\left[-\frac{(\mu-\frac{V_{A}}{\eta v}\pm \frac{1}{\eta xR})^{2}}
{\Delta \mu^{2}}\right],\label{eq:appdpg}
\ena
for the gyro-resonant acceleration $n=\pm 1$, and $d^{3}k=2\pi k^{2}dk d\eta$ has been used
 for fast modes. In the above equation, $L$ is the injection scale of turbulence, $w={k_{\perp}
 v_{\perp}}/{\Omega}, x=k/k_{\min}=kL, R=vk_{\min}/\Omega, M_{\A}^{2}=
 \delta V^{2}/V_{\A}^{2}$, $\eta=\cos\theta$, $k_{c}$ is the cut-off
  of turbulence cascade due to damping, and $J_{n}$ is second order 
 Bessel function.

For the transit time acceleration (TTD), $n=0$, we obtain
\bea
D_{pp}(\mu)^{\rm TTD}=\frac{v\sqrt{\pi}\Omega^{2}(1-\mu^{2})m^{2}V_{\A}^{2}
M_{\A}^{2}}{2LR^{2}}\int_{1}^{k_{c}L}x^{-5/2}dx\int_{0}^{1} \frac{d\eta}{\eta\Delta \mu}J_{1}^{2}(w){\rm exp}
\left[-\frac{(\mu-\frac{V_{\A}}
{\eta v})^{2}}{\Delta \mu^{2}}\right],\label{eq:appdpt}
\ena
where $L$ is the injection scale of turbulence, $w={k_{\perp}
v_{\perp}}/{\Omega}, x=k/k_{\min}=kL, R=vk_{\min}/\Omega, M_{\A}^{2}=
\delta V^{2}/V_{\A}^{2}$.

In the QLT, $R_{n}$ is replaced by $\delta$ function, Equations 
(\ref{eq:appdpg}) and (\ref{eq:appdpt}) become
\bea
D_{pp}(\mu)^{\rm G}=\frac{v\Omega^{2}(1-\mu^{2})m^{2}V_{\A}^{2}M_{\A}^{2}}
{4LR^{2}}\int_{1}^{k_{c}L}x^{-5/2}dx\int_{0}^{1} \frac{d\eta}{\eta}[J_{0}^{2}(w)+J_{2}^{2}(w)]\delta\left(\mu-\frac{V_{A}}{\eta v}\pm \frac{1}{\eta xR}\right),\label{appdpg}
\ena
and 
\bea
D_{pp}^{\TTD}=\frac{v\Omega^{2}(1-\mu^{2})m^{2}V_{\A}^{2}M_{\A}^{2}}
{2LR^{2}}\int_{1}^{k_{c}L}x^{-5/2}dx\int_{0}^{1} \frac{d\eta}{\eta}[J_{1}^{2}(w)]\delta\left(\mu-\frac{V_{A}}{\eta v}\right).
\ena


\begin{thebibliography}{}
\bibitem[]{} Armstrong, J. W., Rickett, B. J., \& Spangler, S. R. 1995, \apj, 443, 209 
\bibitem[]{} Beresnyak, A. 2011, Phys. Rev. Lett., 106, 075001
\bibitem[]{} Beresnyak, A., \& Lazarian, A. 2010, \apj, 722, 110
\bibitem[]{}Chandran, B. D. G., Li, B., Rogers, B. N., Quataert, E., 
\& Germaschewski, K. 2010, \apj, 720, 503

\bibitem{} Chepurnov, A., \& Lazarian, A. 2010, \apj, 710, 853
 
\bibitem[Cho \& Lazarian(2002)]{2002PhRvL..88x5001C} Cho, J., \& Lazarian, 
A. 2002, Phys. Rev. Lett., 88, 245001

\bibitem[]{} Cho, J., Lazarian, A., \& Vishniac, E. T. 2002, \apj, 564, 291

\bibitem[Cho \& Lazarian(2003)]{2003MNRAS.345..325C} Cho, J., \& Lazarian, 
A. 2003, \mnras, 345, 325

\bibitem[]{} Chokshi, A., Tielens, A. G. G. M., Hollenbach, D. 1993, \apj, 407, 806
\bibitem[]{} Dolginov, A. Z., \& Mitrofanov, I. G. 1976, Ap\&SS, 43, 291


\bibitem[Draine(1985)]{1985prpl.conf..621D} Draine, B. T. 1985, Protostars and 
Planets II, 621

\bibitem{} Draine, B. T. 2011, Physics of the Interstellar and Intergalactic Medium 
(Princeton, NJ: Princeton Univ. Press)
\bibitem{} Draine, B.~T., \& Lazarian, A. 1998, \apj, 508, 157
\bibitem[Draine \& Salpeter(1979)]{1979ApJ...231..438D} Draine, B. T., \& 
Salpeter, E. E. 1979, \apj, 231, 438

\bibitem[]{DrSu87} Draine, B. T., \& Sutin, B. 1987, \apj, 320, 803
\bibitem[]{}Draine, B.~T., \& Weingartner, J. 1996, ApJ, 470, 551

\bibitem[]{} Dullemond, C. P., \& Dominik, C. 2005, A\&A, 434, 971
\bibitem[]{} Dung, R., \& Schlickeiser, R. 1990a, A\&A, 237, 504
\bibitem[]{} Dung, R., \& Schlickeiser, R. 1990b, A\&A, 240, 537
\bibitem[]{} Fisk, L. A. 1976, J. Geophys. Res., 81, 4633
\bibitem[]{} Goldreich, P., \& Sridhar, S. 1995, \apj, 438, 763
\bibitem[]{} Hirashita, H., \& Yan, H. 2009, \mnras, 394, 1061 
\bibitem[]{} Hirashita, H., Nozawa, T., Yan, H., \& Kozasa, T. 2010, \mnras, 404, 1437

\bibitem[]{Hoang10} Hoang, T., Draine, B.~T., \& Lazarian, A. 2010, \apj, 715, 1462
\bibitem[]{Hoang11} Hoang, T., \&  Lazarian, A. 2011, \apj, submitted
\bibitem[]{Hoang11} Hoang, T.,  Lazarian, A., \& Draine, B.~T. 2011, \apj, 741, 87
\bibitem[]{}Ivlev, A. V., Lazarian, A., Tsytovich, V. N., de Angelis, U., Hoang, T., Morfill, G. E. 2010, \apj, 723, 612
\bibitem[]{} Jokipii, J. R. 1966, \apj, 146, 480

\bibitem[Jordan 
\& Weingartner(2009)]{2009MNRAS.400..536J} Jordan, M.~E., \& Weingartner, J.~C.\ 2009, \mnras, 400, 536 
\bibitem[]{} Kowal, G., \& Lazarian, A. 2010, \apj, 720, 742
\bibitem{756} Lazarian, A. 2007, J. Quant. Spectrosc. Rad. Trans., 106, 225
\bibitem[Lazarian(1995)]{1995ApJ...453..229L} Lazarian, A.\ 1995, \apj, 
453, 229 
\bibitem[Lazarian 
\& Hoang(2007)]{2007MNRAS.378..910L} Lazarian, A., \& Hoang, T.\ 2007a, \mnras, 378, 910
\bibitem[Lazarian 
\& Hoang(2007)]{2007ApJ...669L..77L} Lazarian, A., \& Hoang, T.\ 2007b, \apjl, 669, L77 
\bibitem[]{} Lazarian, A. 2009, Space Sci. Rev., 143, 357
\bibitem[Lazarian \& Yan(2002)]{2002ApJ...566L.105L} Lazarian, A., \& Yan, 
H. 2002, \apjl, 566, L105
\bibitem[]{} 	Lithwick, Y., \& Goldreich, P. 2001, \apj, 562, 279
\bibitem[]{} Purcell, E.~M. 1969, Physica, 41, 100
\bibitem[]{} Roberge, W. G., Degraff, T. A., \& Flaherty, J. E. 1993, \apj, 418, 287
\bibitem[Roberge 
\& Lazarian(1999)]{1999MNRAS.305..615R} Roberge, W.~G., \& Lazarian, A.\ 1999, \mnras, 305, 615
\bibitem{SM} Schlickeiser, R., \& Miller, J. A. 1998, \apj, 492, 352
\bibitem{SM} Schlickeiser, R. 2002, Cosmic Ray Astrophysics (Berlin: Spinger)
\bibitem{SM}Shalchi, A. 2005, Phys. Plasmas, 12, 052905
\bibitem[]{} V\"{o}lk, H. J. 1975, Rev. Geophys. Space Phys. 13, 547
\bibitem[Yan \& Lazarian(2003)]{2003ApJ...592L..33Y} Yan, H., \& Lazarian, A.
2003, \apjl, 592, L33
\bibitem{Yan04} Yan, H., Lazarian, A., \& Draine, B. T. 2004, \apj, 616, 895 (YLD04)
\bibitem[Yan]{} Yan, H., \& Lazarian, A. 2008, \apj, 673, 942 (YL08)
\bibitem[Yan]{} Yan, H., \& Lazarian, A., \& Petrosian, V. 2008, \apj, 684, 1461 (YLP08)
\bibitem[Yan]{} Yan, H. 2009, \mnras, 397, 1093
\bibitem[]{} Weingartner, J. C.,\& Draine, B. T. 2001, \apjs, 134, 263
\end{thebibliography}
\end{document}